\documentclass[5p,times,final]{elsarticle}

\journal{Parallel Computing}

\usepackage{listings}
\usepackage{booktabs}           %
\usepackage{multirow}           %
\usepackage{subcaption}         %

\begin{document}

\begin{frontmatter}

\title{Asynchronous Runtime with Distributed Manager for Task-based Programming Models}
\author{Jaume Bosch}
\ead{jbosch@bsc.es}
\author{Carlos \'Alvarez}
\author{Daniel Jim\'enez-Gonz\'alez}
\author{Xavier Martorell}
\author{Eduard Ayguad\'e}

\begin{abstract}
Parallel task-based programming models, like OpenMP, allow application developers to easily create a parallel version of their sequential codes.
The standard OpenMP 4.0 introduced the possibility of describing a set of data dependences per task that the runtime uses to order the tasks execution.
This order is calculated using shared graphs, which are updated by all threads in exclusive access using synchronization mechanisms (locks) to ensure the dependence management correctness.
The contention in the access to these structures becomes critical in many-core systems because several threads may be wasting computation resources waiting their turn.

This paper proposes an asynchronous management of the runtime structures, like task dependence graphs, suitable for task-based programming model runtimes.
In such organization, the threads request actions to the runtime instead of doing them directly.
The requests are then handled by a distributed runtime manager (DDAST) which does not require dedicated resources.
Instead, the manager uses the idle threads to modify the runtime structures.
The paper also presents an implementation, analysis and performance evaluation of such runtime organization.
The performance results show that the proposed asynchronous organization outperforms the speedup obtained by the original runtime for different benchmarks and different many-core architectures.

\end{abstract}


\end{frontmatter}

\begin{center}
\begin{tabular}{|p{0.9\linewidth}|}
\hline\\
\textbf{LICENCE}\\
$\copyright$ 2020 Elsevier. This manuscript version is made available under the CC-BY-NC-ND 4.0 license \\http://creativecommons.org/licenses/by-nc-nd/4.0/
\\
\\
DOI: 10.1016/j.parco.2020.102664.
\\\\\hline
\end{tabular}
\end{center}

\section{Introduction} \label{sec:1}
The multicore processors popularization started due to the end of Dennard scaling law which states that the power density of an integrated circuit can stay constant meanwhile the transistors get smaller.
Until 2006, Dennard's law and Moore's law have guided processor manufacturers to periodically reduce the transistors length and increase the clock frequency which also increases the processors performance.
However, the leakage current grows much faster at small transistor sizes; therefore the clock frequency cannot increase without impacting the overall power consumption.
Since the transistor still reduces its size periodically as Moore's law states, processor manufacturers started to introduce multiple cores in their processors to keep the processors performance increase.

As multicore processors have become popular, parallel programming has become a need to take advantage of these processors.
Instead of dealing with complex applications programmed for one specific processor architecture, parallel programming models decouple applications from hardware.
Their goal is to allow programmers to indicate the potential parallelism in the applications source code without directly managing it.
There are several examples like MapReduce \cite{mapreduce}, OpenMP \cite{omp}, OpenCL \cite{opencl}, StarSs \cite{starss}, etc.
The exposed parallelism is then managed by a runtime library that coordinates the application execution transparently to the application programmer.
Similarly, there are parallelized libraries that can be used from sequential applications that implement several parallel skeletons for commonly used operations.
Some examples of these libraries are: Spark \cite{spark}, OpenBLAS \cite{openblas}, Intel MKL, etc.

The task oriented paradigm is one powerful way to define potential parallelism in one application.
Programmers only have to annotate code regions called tasks that can run in parallel.
Additionally, developers can provide additional task information like data requirements. This information defines the task execution order enforced by the runtime libraries at execution time.
The OpenMP standard introduced task dependences in the 4.0 version greatly influenced by the OmpSs programming model which extends the standard syntax with additional features.

The runtimes of these models are responsible for guaranteeing the task execution order correctness defined by the task data requirements.
Therefore, the runtime updates a task graph when a task is created and when a task finalizes its execution.
Usually, these modifications require to read and write the information in the task graph atomically to ensure the order correctness.

In a processor with a lot of cores, the probability of collisions between threads trying to access the task dependence graph increases.
Each collision implies that a thread is wasting its computation time waiting for another one modifications.
This problem that has currently started arising is expected to be an important bottleneck as the number of cores in the future processors is expected to keep growing \cite{dast}.
Thereby, the access contention on some runtime structures will kill the application performance if runtimes do not redesign their internals to tackle the problem.

To improve the current task-based parallel programming runtimes and avoid the contention expected in the many-core processors, we propose an asynchronous runtime organization \linebreak[4] where the runtime threads do not update the runtime structures directly.
Instead, the threads request the needed actions to the runtime and this request will be handled in the future.
This asynchronous approach avoids the problem of actively waiting for the exclusive access and allows the threads to return immediately to the application code.
Moreover, such structure tries to maximize the utilization of the processor cores to run application code and avoid active waiting on the locks.

The threads requests to the runtime are handled by a runtime manager who updates the runtime structures.
Initially, we proposed a centralized implementation based on an extra thread (DAS Thread, DAST) together with a mechanism to avoid the manager saturation \cite{dast}.
In this work, we present a new distributed implementation of the runtime manager (Distributed DAST, DDAST) based on a mechanism where any thread may become a runtime manager thread.
Therefore, the runtime tries to use all the available threads in a smart way to restrict the accesses to the runtime structures.

The proposed asynchronous runtime model provides similar performance to the original runtime when the application has a small number of tasks or when the execution uses a reduced amount of threads.
However, when the number of tasks and/or the number of threads is large, the new runtime achieves better performance due to the better thread utilization, data locality and contention reduction.
All the changes proposed here are transparent for application developers and are general enough to be used in a wide range of task-based parallel programming models.
Even more, the design could be adapted for particular heterogeneous architectures, like big.LITTLE \cite{big.little}, allowing a subset of the worker threads to become manager threads.

The remainder of the paper is organized as follows.
Section~\ref{sec:2} describes the OmpSs task-based programming model, as an example of task-based programming model whose runtime takes care of the dependences management.
Section~\ref{sec:3} describes the design and implementation of the new distributed runtime manager (DDAST).
Section~\ref{sec:4} presents the experimental setup.
Section~\ref{sec:5} presents the tuning results for the manager internal parameters.
Section~\ref{sec:6} shows the performance of the new runtime and analyzes its behavior during the executions.
Finally, section~\ref{sec:7} presents the related work and section~\ref{sec:8} concludes.

\section{Background} \label{sec:2}
The implementation of the asynchronous runtime model has been developed using the OmpSs programming model, which is a forerunner of the standard OpenMP parallel programming model.
Therefore, the following sections introduce the OmpSs programming model (section~\ref{sec:2:ompss}) and Nanos++ (section~\ref{sec:2:nanox}), which is the runtime library used to run OmpSs applications.
The OmpSs programming model is also supported by the source-to-source Mercurium compiler for C, C++ and Fortran.
However, the change proposed in this work is only to modify the runtime internals which is transparent for the compiler and the application developers.
Besides, the design fits any runtime for task-based parallel programming models because they share the same execution model.

\subsection{OmpSs programming model basics} \label{sec:2:ompss}

The OmpSs programming model is a task-based parallel programming model developed at BSC and composed by a set of directives and library routines.
The name OmpSs comes from two other programming model names, OpenMP \cite{omp} and StarSs \cite{starss}.
The goal of the programming model was to extend the OpenMP syntax with some of the StarSs features to provide a productive environment for HPC applications development.
Productive means that the applications developed in OmpSs achieve reasonable performance compared to similar solutions for the same architectures and that the development cost is small and does not require huge changes in the applications \cite{ompss_paper} \cite{bueno2015run}.

On one hand, OmpSs takes from OpenMP the philosophy of providing a way to produce a parallel version of the application adding annotations that do not require modifications in the source code.
These annotations allow the compiler to generate a parallel version of the application by replacing the annotations with runtime API calls.
This philosophy is intended to simplify the development process leading to a better productivity.
On the other hand, OmpSs takes from StarSs the thread-pool model.
In contrast to the fork-join model used by OpenMP, StarSs model has an implicit parallelism during all the execution so, programmers do not have to annotate the parallel regions \cite{ompss_paper} \cite{bueno2015run}.

\subsubsection{Task annotation}

\vspace{\parskip}
The main annotation in OmpSs is the task clause which defines a code region that will be asynchronously executed.
Tasks can be concurrently executed by any thread when they are ready.
The task execution order can be defined by the programmers using the \texttt{in(...)}, \texttt{out(...)} and \texttt{inout(...)} clauses that extend the task annotation.
These clauses define the input, output and input+output data dependences for each task (similar to OpenMP \texttt{depend} clause), which implicitly define the dependences between tasks.
The runtime is responsible for synchronizing task executions to guarantee the data dependences.

In addition to the implicit synchronization created by the data dependences, the programmer can introduce explicit synchronization points using the taskwait annotation.
This ensures that after the annotation all tasks created before it are executed and all data generated by tasks is available with the latest values. \newline

\includegraphics[width=.9\linewidth]{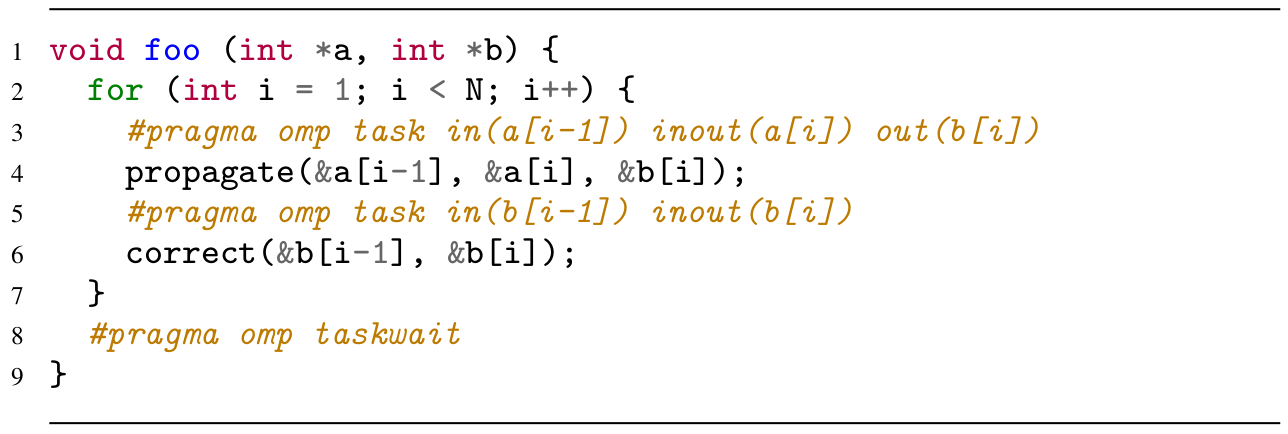}%
\begin{lstlisting}[caption={OmpSs code annotation example}, label={lis:2:ex_src}]
\end{lstlisting}

\begin{figure}[ht]
  \centering
  \includegraphics[width=1.\linewidth]{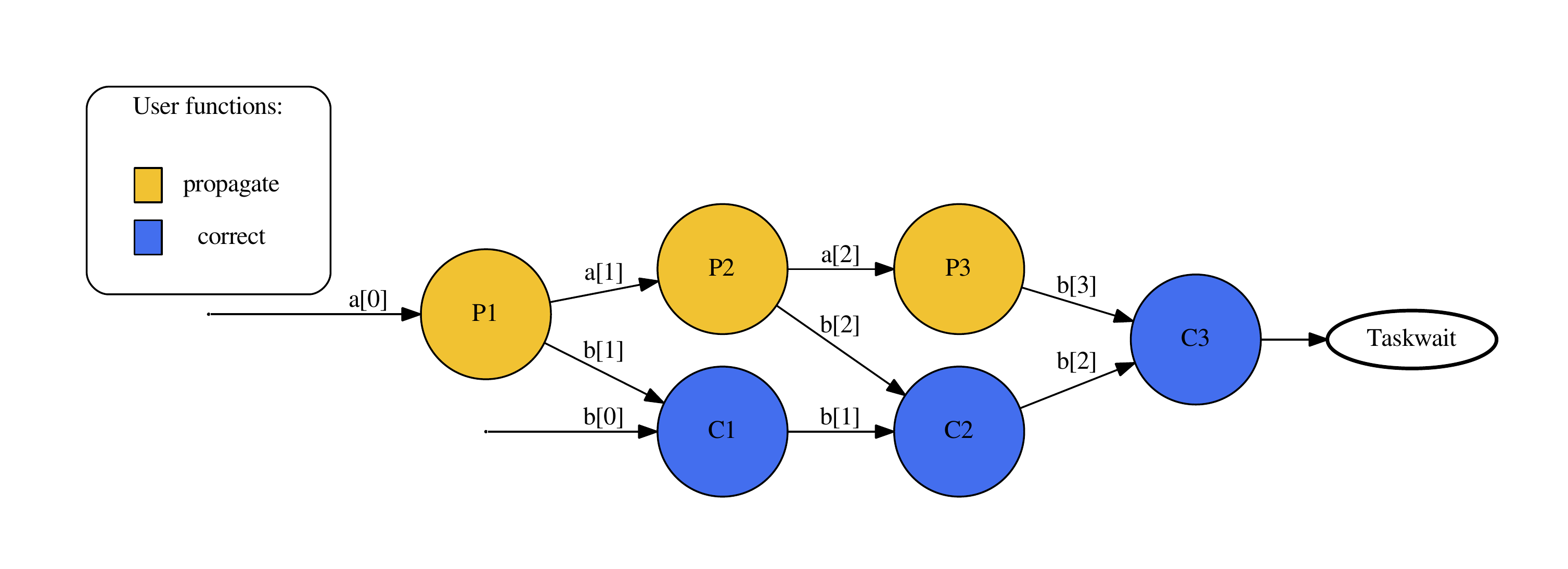}
  \captionof{figure}{OmpSs task graph for listing~\ref{lis:2:ex_src} (N=3)}
  \label{fig:2:ex_graph}
\end{figure}

An example code \cite{ompss:user_guide} of a C function parallelized with OmpSs is shown in listing~\ref{lis:2:ex_src}.
The function contains two function calls that are annotated with the task directive: \textit{propagate} and \textit{correct}.
Thus, the calls to those functions are asynchronously executed when the data dependences defined in the task annotation are satisfied.
The resulting task dependence graph (for $N=3$) is shown in figure~\ref{fig:2:ex_graph} where the nodes are tasks and the edges true dependences among them.
There is a true dependence between the $i^{th}$ \textit{propagate} and the $i^{th}$ \textit{correct} tasks due to the $i^{th}$ element of \textit{b}.
There is a true dependence between the $i^{th}$ \textit{propagate} and the $i^{th}+1$ \textit{propagate} tasks due to the $i^{th}$ element of \textit{a}.
Finally, there is a true dependence between the $i^{th}$ \textit{correct} and the $i^{th}+1$ \textit{correct} tasks due to the $i^{th}$ element of \textit{b}.

\subsection{Nanos++ runtime} \label{sec:2:nanox}

Nanos++ is a runtime library designed to serve as runtime support in parallel environments.
The runtime is developed at the Barcelona Supercomputing Center within the Programming Models group, and its main target is to support the OmpSs programming model.
Apart from OmpSs, Nanos++ also supports most of the OpenMP 3.1 features and includes some additional extensions (some of them also introduced in following OpenMP releases) \cite{nanox}.

The runtime provides the required services to support task parallelism based on data dependences.
Data parallelism is also supported by means of services mapped on top of its task support.
Tasks are implemented as user-level threads when possible.
It also provides support for maintaining coherence across different address spaces (such as nodes with FPGAs or cluster nodes) by means of a directory/cache mechanism \cite{nanox}.

The main purpose of Nanos++ is to be used in research of parallel programming environments.
The runtime tries to enable easy development of different parts, so researchers have a platform that allows them to try different mechanisms.
As such it is designed to be extensible by means of plugins.
The scheduling policy, the throttling policy, the dependence \linebreak[4] approach, the barrier implementations, slicers and worksharing mechanisms, the instrumentation layer and the architectural dependant level are examples of plugins that developers may easily implement using Nanos++ \cite{nanox}.

\subsubsection{Task life cycle}

Task representation inside Nanos++ is made by one Work Descriptor (WD) for each task.
Each WD contains all needed information to manage the task during its life cycle.
For instance, the WDs store the data dependences of each task.
The parent task, which is the task being run when the child task is created, contains the task graph with the relations of its children.
This limits the tasks to depend on only sibling tasks, but the global order is guaranteed because parent task dependences must be a super-set of its child tasks dependences.
Despite this distributed model, actions in each graph are protected by spin-locks because different sibling tasks can finalize at the same time and/or collision with another sibling task creation.

The different steps in the task life cycle are summarized following: %
\begin{enumerate}
   \item Task creation.
   At this step, the WD structure is allocated and initialized with the information provided in the annotations related to the task.
   Moreover, the values of function arguments or local variables are stored in order to execute the code asynchronously.

   \item Task submission.
   At this step, the data dependences of the task are stored in the WD and introduced in the task graph to compute the predecessor WDs.
   If no predecessors are found, the task can immediately become ready.
   How the predecessors are computed depends on the dependences plugin which can be changed in each execution.

   \item Task becomes ready.
   At this step, task data dependences have been satisfied or task blocking condition has been fulfilled.
   Consequently, the task execution can start.
   How the task will be executed depends on the used scheduling policy that can be changed in each execution.

   \item Task becomes blocked.
   At this step, the task cannot proceed its execution until some condition becomes true.
   For example, when a task contains a \texttt{taskwait} annotation it becomes blocked until its children tasks finish.

   \item Task finalization.
   At this step, the task has finished its execution and the successor WDs may become ready if they only depend on the finalized task.
   Therefore, the WD can be deleted if it does not have children tasks.
   Otherwise, the children tasks might further reference the parent WD in their finalization to access the task graph.

   \item Task deletion.
   At this step, the WD can be safely deleted because no more references to it will be done.
\end{enumerate}

\begin{figure}[htp]
  \centering
  \includegraphics[width=0.85\linewidth]{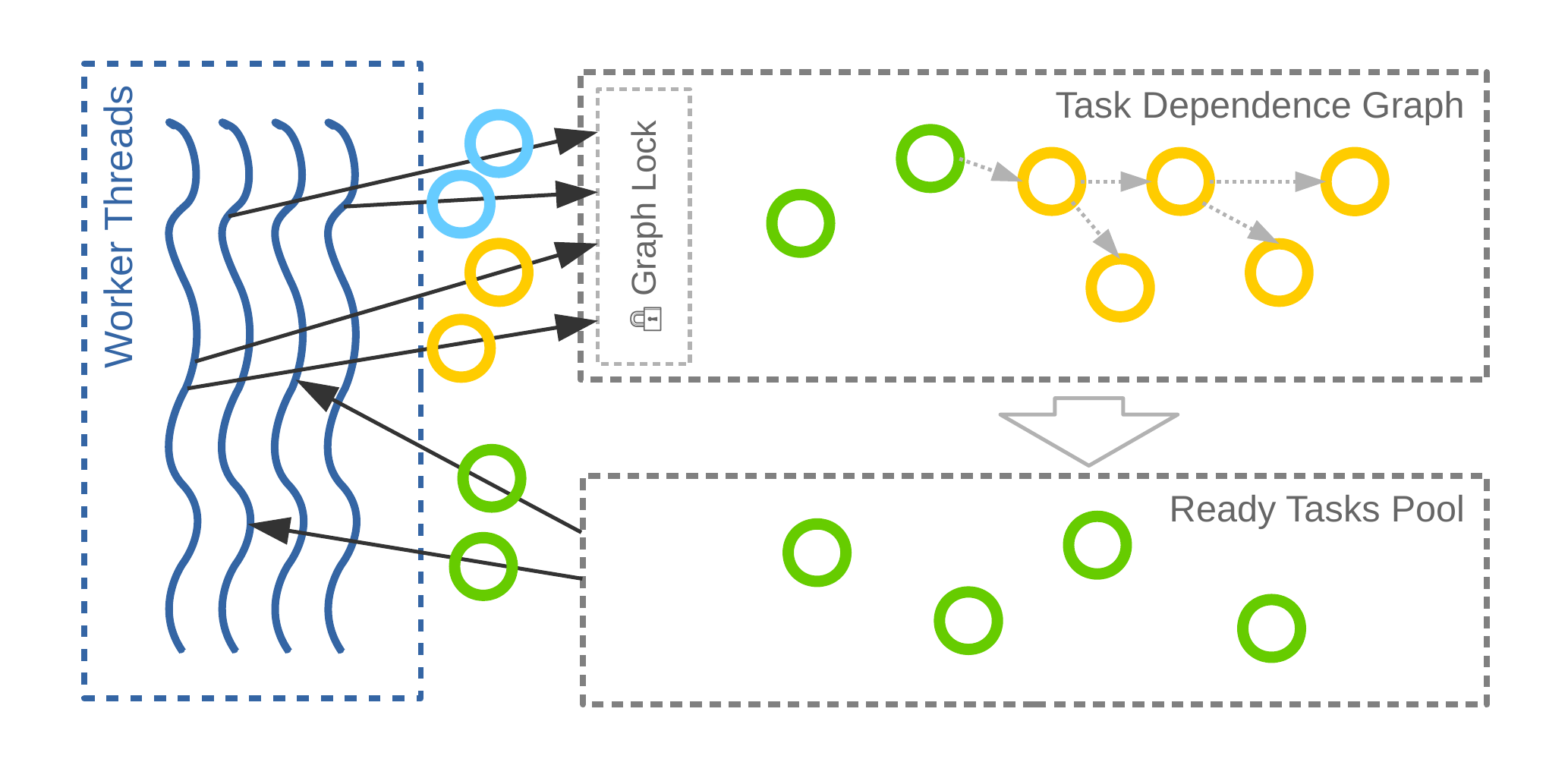}
  \caption{Nanos++ task flow over runtime structures}
  \label{fig:2:task_flow}
\end{figure}

Each task state is mainly related with one runtime component, therefore the WDs flow over the different runtime structures during its life cycle.
Figure~\ref{fig:2:task_flow} shows a scheme of the tasks flow where each circle represents a task.
Each circle color is associated with a task state: yellow for a task being created or a submitted task, green for a ready task and blue for a finished task.
First, a thread pushes the created tasks into the task dependence graph to determine the task order.
Then, other threads ``push'' the finalized tasks into the task dependence graph to notify the successor tasks.
In addition, this action removes the finished task from the graph and adds the tasks that become ready into the ready tasks pool.
Both dependence graph operations (push a created or finished task) require acquisition of the graph lock to safely perform the modifications.
Finally, the worker threads try to acquire ready tasks from the ready tasks pool to execute them.
The management (insertion, deletion, etc.) of tasks into the ready tasks pool may require acquiring some lock, but it is not shown on the figure~\ref{fig:2:task_flow} as the pool implementation depends on the scheduling policy.

\section{Distributed Runtime Manager} \label{sec:3}
\begin{figure}[htp]
  \centering
  \includegraphics[width=0.85\linewidth]{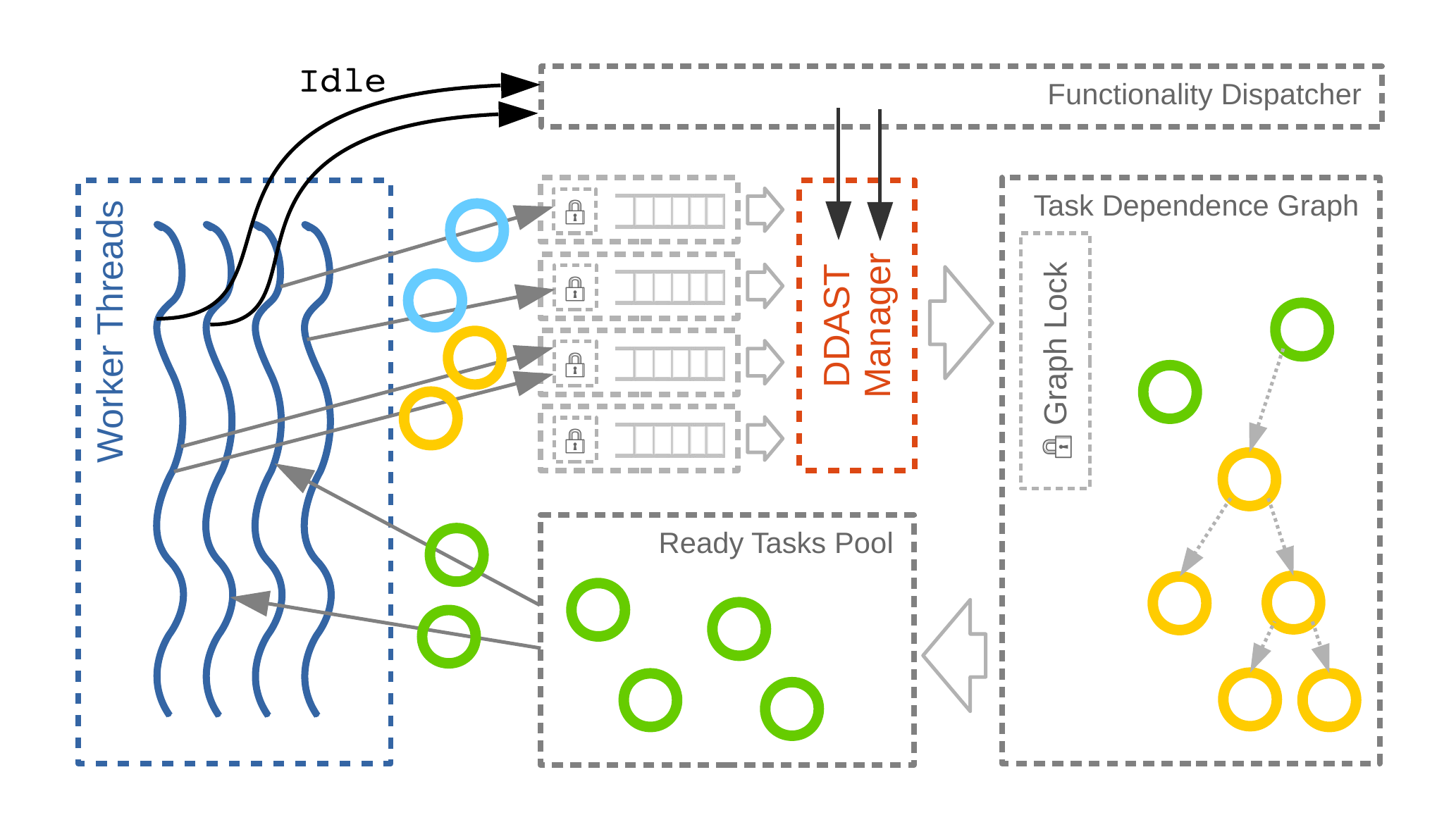}
  \caption{Task flow over runtime structures for the asynchronous runtime with DDAST manager}
  \label{fig:3:task_flow}
\end{figure}

The design of the asynchronous runtime with the distributed manager is based on the idea that any worker thread can become a manager thread and start executing only runtime code.
With this approach, all threads can cooperate to execute the pending runtime operations when there are several of them.
Correspondingly, all the threads can execute application tasks when the number of pending runtime operations is small.
The implementation of the newer design is based on the knowledge acquired in our previous design where almost all the runtime operations that modify the runtime structures were centralized in an additional thread called DAST (DAS Thread)\cite{dast}.
However, the development of the new distributed runtime manager (DDAST, Distributed DAST) is done over a new and fresh runtime version and it is based on general modules that can be extended to support other runtime functionalities.
The three main parts of the new asynchronous runtime design with the DDAST manager can be seen in figure~\ref{fig:3:task_flow} and are explained as follows:

\begin{itemize}
   \item Section~\ref{sec:3:msgs} explains the messages (requests of runtime operations) that the worker threads send to the DDAST manager, instead of directly doing those operations, and the queues used to transmit/store them.
   \item Section~\ref{sec:3:dispch} explains the Functionality Dispatcher module introduced in the runtime to mediate between its components and used by the DDAST manager.
   \item Section~\ref{sec:3:callbk} explains the module that implements the \linebreak[4] DDAST callback registered into the Functionality Dispatcher.
\end{itemize}

\subsection{Messages and Queues} \label{sec:3:msgs}

The messages (request of runtime operations) sent by the worker threads to the runtime manager can be of two types:
the \textit{Submit Task Message} and the \textit{Done Task Message}.
The first one, the \textit{Submit Task Message} is sent when a worker thread wants to submit a new task into the runtime structures to find out its predecessor tasks.
The second one, the \textit{Done Task Message} is sent when a worker thread finishes the execution of a task and wants to notify the successor tasks, scheduling them if they become ready.

There is another synchronization point between the worker threads and the runtime manager that could be managed with a third message.
That is during the task deletion step when the task information is deleted.
At that point, the worker threads do not know if the task \textit{Done Task Message} has been handled or not, so they cannot safely delete the task information.
However, based on our previous experience \cite{dast}, this synchronization can be handled by means of an additional task state.
This allows to safely remove the task information with less overhead than using a third message to the runtime manager.

The two message types are stored in a queuing system until the runtime manager does not handle them.
There are two independent queues (one for each message type) replicated for each worker thread where only itself can insert messages, and only the DDAST manager can pop messages, as shown in figure~\ref{fig:3:task_flow}.
It is important to stress that the queue for the \textit{Submit Task Message}s respects the insertion order to create the right task dependence graph.
Otherwise, the sequential execution order of the tasks cannot be guaranteed.
Also, only one manager thread can pop and process \textit{Submit Task Message}s from a worker thread queue at the same time.
Otherwise, a newer message could enter in the task dependence graph, before an older one, creating a wrongly computed task dependence graph.
In contrast, the \textit{Done Task Message}s can be processed by any manager thread concurrently without any restriction, as there is no implicit guaranteed finalization order for the tasks under execution.

\subsection{Functionality Dispatcher} \label{sec:3:dispch}

The Functionality Dispatcher is a new module introduced in the runtime core that mediates between different runtime parts.
This module easily allows both using the idle resources to execute any runtime operation and implementing some runtime functionalities without having computational resources exclusively dedicated to them.

Any runtime module can register a callback function in the Functionality Dispatcher during the runtime initialization or the application execution.
Those callbacks are listed into the new module, which is also notified by the worker threads when they become idle.
In current Nanos++ implementation, the worker threads make a busy waiting loop until they obtain tasks to execute.
Therefore, the Functionality Dispatcher tries to take advantage of those idle resources and uses them to execute the different registered callbacks.

\begin{figure}[ht]
  \centering
  \includegraphics[width=0.98\linewidth]{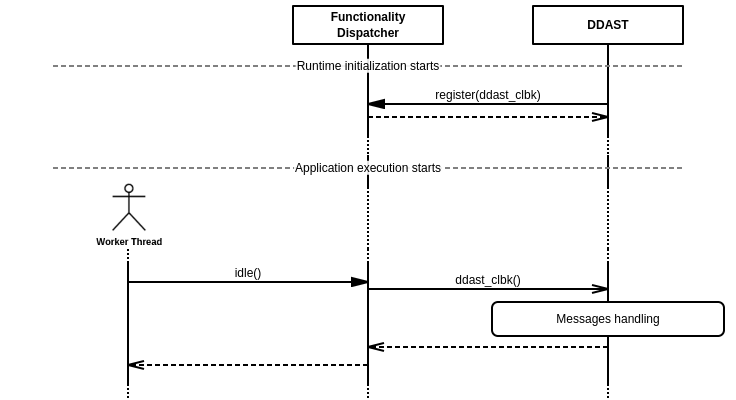}
  \caption{Functionality Dispatcher Sequence Diagram Example}
  \label{fig:3:dispch}
\end{figure}

\hyphenation{re-gis-ters}
Figure~\ref{fig:3:dispch} shows the sequence diagram for the implemented Functionality Dispatcher with the DDAST manager.
During the runtime initialization, the DDAST module registers a callback function into the Functionality Dispatcher.
Therefore, during the application execution, a worker thread that becomes idle notifies the Functionality Dispatcher, which instructs the worker to execute the DDAST callback function that starts handling the messages in the queuing system.
This final change of a worker thread to a manger thread through the Functionality Dispatcher is also shown in figure~\ref{fig:3:task_flow}.

\subsection{DDAST Callback} \label{sec:3:callbk}

The distributed runtime manager is implemented in a callback function registered in the Functionality Dispatcher.
Therefore, the callback is executed when a worker thread becomes idle and the Functionality Dispatcher calls the registered function.
That a worker thread becomes idle usually means that the pending messages in the queues must be processed to submit more tasks into the task graph or trigger the scheduling of some new ready tasks.

The behavior of the DDAST callback is parametrized by different constants defined at the beginning of the application execution.
The performance impact and the default values for these variables are analyzed in section~\ref{sec:5}.
Here follows a brief list and explanation of these variables::

\begin{itemize}
   \item \texttt{MAX\_DDAST\_THREADS}.
   Maximum number of threads allowed to execute the DDAST callback concurrently.

   \item \texttt{MAX\_SPINS}.
   Number of times that the thread will try to get messages without success before leaving the callback.

   \item \texttt{MAX\_OPS\_THREAD}.
   Maximum number of messages satisfied from the same worker thread queue before changing to another worker thread queue.

   \item \texttt{MIN\_READY\_TASKS}.
   Minimum number of ready tasks available before exiting exit the callback.
\end{itemize}

\includegraphics[width=.9\linewidth]{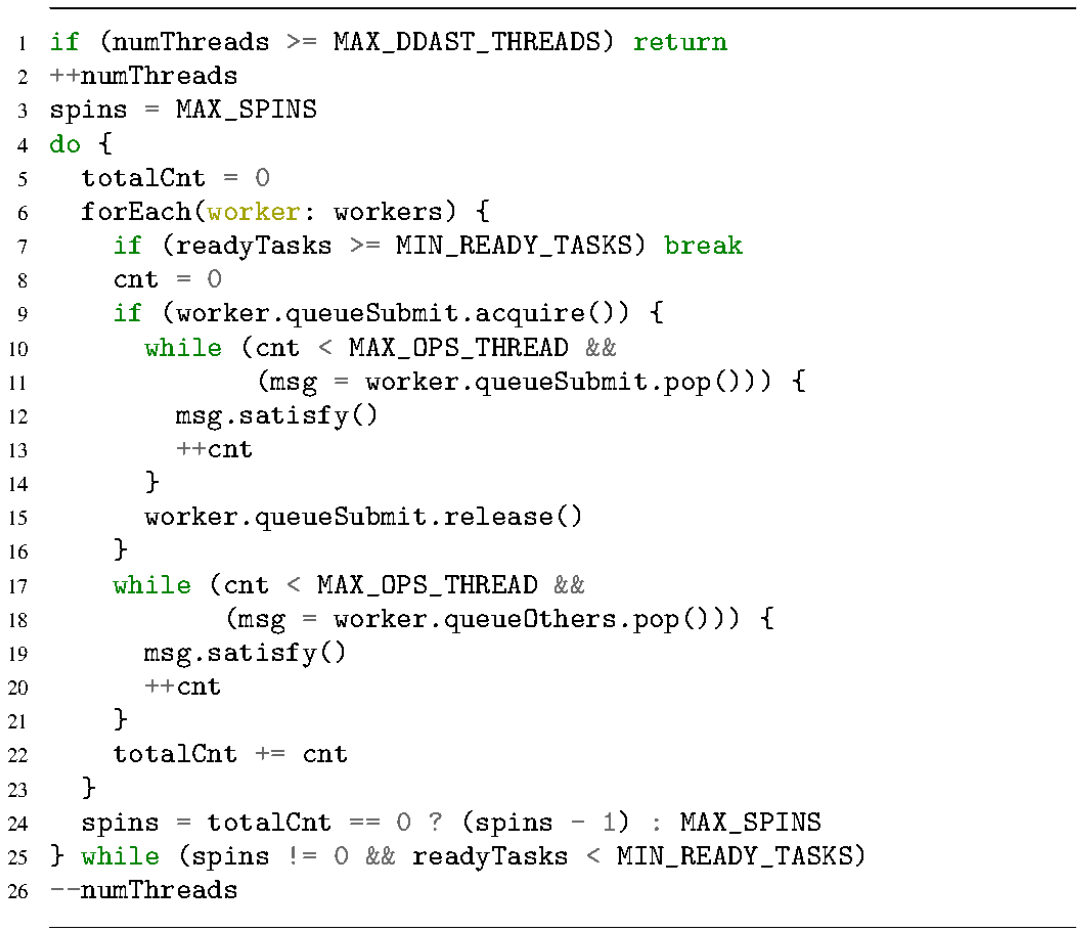}%
\begin{lstlisting}[caption={DDAST callback pseudo-code}, label={lis:3:callbk}]
\end{lstlisting}

Listing~\ref{lis:3:callbk} shows the pseudo-code of the callback function.
First, the number of threads running the DDAST callback is checked and the function returns if the maximum number is reached (listing~\ref{lis:3:callbk}, line 1).
After that, the idle thread tries to retrieve messages and satisfy them.
This is done until the minimum number of ready tasks is reached or the thread iterates \texttt{MAX\_SPINS} times without finding any message (listing~\ref{lis:3:callbk}, line 25).
The way to retrieve messages is to iterate through all worker threads and try to retrieve and satisfy up to \linebreak[4] \texttt{MAX\_OPS\_THREAD} messages for the same worker thread (prioritazing the \textit{Submit Task Message}s).
Note that not all worker threads are iterated if the number of ready tasks becomes higher than \texttt{MIN\_READY\_TASKS} (listing~\ref{lis:3:callbk}, line 7).

\section{Experimental Setup} \label{sec:4}
The evaluation of the asynchronous runtime implementation has been done in different architectures and under different benchmarks to show the adaptability and capabilities in different contexts.
They are introduced and explained in the following sections: section~\ref{sec:4:env} for the different machines/architectures and section~\ref{sec:4:bench} for the different benchmarks.
However, a common criteria for all the architectures and benchmarks has been followed to enhance the evaluation quality, avoid external inferences and facilitate the reproducibility of the results.

\begin{enumerate}
   \item The machine nodes used during all executions are exclusively reserved for the tests.
   \item The applications are compiled with optimization flags and tuning for each architecture.
   \item The shown time measurements are the best execution time of 5 repetitions.
   This way the most stressful conditions, where runtime overheads become crucial, are analyzed.
   However, the variability between executions is small and the results will be qualitatively the same using the average execution time.
   \item The scheduling policy used in all the OmpSs executions is the Distributed Breadth First (DBF).
   The DBF policy uses a queue of ready tasks for each thread with a stealing mechanism \cite{ompss:user_guide}.
\end{enumerate}

\subsection{Machines/Architectures} \label{sec:4:env}

The tests have been run in different processor architectures summarized in table~\ref{tab:4:machines}, the characteristics and software stack of each machine is explained in the next subsections.

\begin{table}[ht]
  \caption{Machine resources summary} \label{tab:4:machines}
  \centering
  \small
\begin{tabular}{@{}lccccc@{}}
\toprule
\multirow{2}{*}{\textbf{Machine}} & \textbf{Num.}  & \textbf{Threads} & \textbf{CPU} & \textbf{Mem.} & \multirow{2}{*}{\textbf{Other}} \\
                                  & \textbf{Cores} & \textbf{x core}  & \textbf{Ghz} & \textbf{GB} & \\ \midrule
KNL          & 64     & 4                & 1.3     & 96      & 16GB HBM \\
ThunderX     & 48     & 1                & 1.8     & 64      & \\
Power8+      & 10+10  & 8                & 4       & 256     & \\
Power9       & 20+20  & 4                & 3       & 512     & \\ \bottomrule
\end{tabular}
\end{table}

\subsubsection{Intel Xeon Phi (KNL)}

The Knights Landing (KNL) is the second generation of a series of processors manufactured by Intel and characterized by its high parallelism and vectorization capacity \cite{knl}.
The model used in our evaluation is the \textit{Intel{\textregistered} Xeon{\textregistered} Phi(TM) CPU 7230} configured in Quadrant mode \cite{knl}.
The executions use up to 64 worker threads, 1 thread per core, as the used machine has the hyper-threading disabled.
In the executions with less than 64 cores, the use of the processor resources is maximized.
For instance, the cores 0 and 2 are used in a 2-cores execution because cores 0 and 1 share part of the cache.

The version of the Intel{\textregistered} Math Kernel Library (Intel(R) MKL) used by the applications is 2017.0.2.
The compiler used to compile the applications and the runtimes is the GNU C Compiler Collection (GCC) version 6.3.0.

\subsubsection{ThunderX (ARM)}

The ThunderX is a family of processors developed by Cavium based on the 64-bit ARMv8 architecture \cite{thunderx}.
The compiler used to natively compile the applications and the runtimes is the GNU C Compiler Collection (GCC) version 5.3.0.
The version of the ARM Performance Libraries (ARM PL) used by the applications is 2.0.0.

\subsubsection{Power8+}

The Power8+ is a revision of the Power8 processors developed by IBM.
The nodes used in our evaluation have 2 IBM PowerNV 8335-GTB processors with 10 cores each.
The executions only use 1 and 2 threads per core because more than that does not benefit the performance of the evaluated benchmarks.
The compiler used to natively compile the applications and the runtimes is the GNU C Compiler Collection (GCC) version 6.3.0.

\subsubsection{Power9}

The Power9 is the latest version of the architecture developed by IBM \cite{power9}.
The nodes used in our evaluation have 2 IBM Power9 8335-GTG processors with 20 cores each.
The executions only use 1 thread per core because more than that does not benefit the performance of the evaluated benchmarks.
The compiler used to natively compile the applications and the runtimes is the GNU C Compiler Collection (GCC) version 8.1.0.

\subsection{Benchmarks} \label{sec:4:bench}

The used benchmarks are explained in the following subsections.
For each one, its execution arguments are explained and provided with the number of created tasks in each configuration and any other remarks that may be valuable for reproducibility.
In all of them, some timing instructions are added after the sequential initialization and after the final global taskwait.
The elapsed time between these two points is defined as the execution time in the rest of the paper.

For each benchmark, two different sets of execution parameters are used to create two tasks granularities: coarse grain (CG) tasks and fine grain (FG) tasks.
Besides, the benchmark execution parameters are selected considering the following:
\begin{itemize}
   \item Problem size. Have a big enough problem size to gather significant results.
   \item Task size (CG). Smallest task size that has enough parallelism to feed all processor cores, delivering almost the best performance hiding the Nanos++ runtime overheads.
   \item Task size (FG). Solve the same problem with tasks that use half the coarse grain value.
\end{itemize}

\subsubsection{Matrix Multiply}

The Matrix Multiply (Matmul) benchmark \cite{bench:bar} computes the product of two blocked matrices in parallel.
The application takes two main arguments: the matrix dimension (\texttt{MS}) and the block dimension (\texttt{BS}).
The task dependences follow a regular pattern with several independent chains that group all tasks working with the same output block.
The used values for \texttt{MS} and \texttt{BS} arguments are summarized in table~\ref{tab:4:dmm}.

\begin{table}[ht]
  \caption{Matmul execution arguments} \label{tab:4:dmm}
  \centering
  \small
\begin{tabular}{@{}lccccc@{}}
\toprule
\multirow{2}{*}{\textbf{Machine}} & \multirow{2}{*}{\textbf{MS}} & \multicolumn{2}{c}{\textbf{Coarse Grain}} & \multicolumn{2}{c}{\textbf{Fine Grain}}   \\
                                  &                                       & \textbf{BS} & \textbf{\#Tasks} & \textbf{BS} & \textbf{\#Tasks} \\ \midrule
KNL                               & 8.192                                  & 512                 & 4.096                & 256                 & 32.768               \\
ThunderX                          & 4.096                                  & 128                 & 32.768               & 64                  & 262.144              \\
Power8+/9                         & 8.192                                  & 512                 & 4.096                & 256                 & 32.768               \\ \bottomrule
\end{tabular}
\end{table}

\subsubsection{N-Body}

The N-Body benchmark \cite{bench:nbody} simulates movements of particles under some physic forces, such as gravity.
The application takes three arguments: the number of particles \linebreak[4] (\texttt{NUM\_PARTICLES}), the number of time steps (\texttt{NUM\_TIMESTEPS}) to be simulated and the number of particles per block (\texttt{BS}).
Therefore, the particles are spread into blocks with \texttt{BS} particles, which are used as task input/output.
The tasks follow a regular chained pattern similar to the Matmul one but this benchmark has nested tasks.
This nesting make more critical some of the requests to the DDAST manager because they may block the application parallelism until they are processed.
The values for the arguments used in each machine are summarized in table~\ref{tab:4:nbody} with the number of tasks created in each configuration.

\begin{table}[ht]
  \caption{N-Body execution arguments} \label{tab:4:nbody}
  \centering
  \small
  \renewcommand\tabcolsep{3pt}
\begin{tabular}{@{}lcccccc@{}}
\toprule
\multirow{2}{*}{\textbf{Machine}} & \textbf{Num.}       & \textbf{Num.}      & \multicolumn{2}{c}{\textbf{Coarse Grain}} & \multicolumn{2}{c}{\textbf{Fine Grain}}   \\
                                  & \textbf{Particles}  & \textbf{Timesteps} & \textbf{BS} & \textbf{\#Tasks} & \textbf{BS} & \textbf{\#Tasks} \\ \midrule
KNL                               & 16.384               & 16                 & 128                 & 262.176              & 64                  & 1.048.608             \\
ThunderX                          & 16.384               & 16                 & 128                 & 262.176              & 64                  & 1.048.608             \\
Power8+/9                         & 16.384               & 16                 & 256                 & 65.568               & 128                 & 262.176               \\ \bottomrule
\end{tabular}
\end{table}

\subsubsection{Sparse LU}

The Sparse LU benchmark \cite{bench:bar} computes the Lower Upper (LU) decomposition of a sparse matrix in parallel.
The application takes two arguments: the matrix dimension (\texttt{MS}) and the block dimension (\texttt{BS}).
Therefore, the matrix with \texttt{MS}$*$\texttt{MS} elements is divided into sub-matrices with \texttt{BS}$*$\texttt{BS} elements.
The task dependences follow a much more complex and irregular pattern than the Matmul and N-Body benchmarks.
The used values for \texttt{MS} and \texttt{BS} arguments are summarized in table~\ref{tab:4:sparselu}.

\begin{table}[ht]
  \caption{Sparse LU execution arguments} \label{tab:4:sparselu}
  \centering
  \small
\begin{tabular}{@{}lccccc@{}}
\toprule
\multirow{2}{*}{\textbf{Machine}} & \multirow{2}{*}{\textbf{MS}} & \multicolumn{2}{c}{\textbf{Coarse Grain}} & \multicolumn{2}{c}{\textbf{Fine Grain}}   \\
                                  &                                       & \textbf{BS} & \textbf{\#Tasks} & \textbf{BS} & \textbf{\#Tasks} \\ \midrule
KNL                               & 8.192                                  & 128                 & 11.472               & 64                  & 89.504               \\
ThunderX                          & 8.192                                  & 128                 & 11.472               & 64                  & 89.504               \\
Power8+/9                         & 8.192                                  & 128                 & 11.472               & 64                  & 89.504               \\ \bottomrule
\end{tabular}
\end{table}

\section{DDAST Tuning} \label{sec:5}
The initial executions with the new runtime structure were intended to find good default values for the callback parameters explained in section~\ref{sec:3:callbk}.
To this end, some initial values are defined, based on a reasonable approximation to the expected optimal ones, and the same execution is repeated changing only one parameter value.
The executions for each parameter are done with two benchmarks that have different task dependence patterns: Matmul and Sparse LU.
Each execution set is doubling the parameter value from 1 up to 128.
Also, different amounts of threads and threads per core are considered depending on the architectures: KNL, ThunderX and Power8+.
However, the results only consider the two configurations with the largest amount of threads in each architecture as they are the most interesting.
Moreover, the larger the number of threads, the bigger the influence of parameter modifications in the execution time.

\begin{table}[ht]
  \caption{DDAST parameters values} \label{tab:5:vals}
  \centering
  \small
\begin{tabular}{@{}lcc@{}}
\toprule
\textbf{Parameter}  & \textbf{Initial Value} & \textbf{Tuned Value} \\ \midrule
\texttt{MAX\_DDAST\_THREADS} & $\infty$     & $\lceil num\_threads/8 \rceil$ \\
\texttt{MAX\_SPINS}          & 20           & 1                    \\
\texttt{MAX\_OPS\_THREAD}    & 6            & 8                    \\
\texttt{MIN\_READY\_TASKS}   & 4            & 4                    \\ \bottomrule
\end{tabular}
\end{table}

The predefined values for each parameter before (\textit{Initial value}) and after (\textit{Tuned Value}) the tuning are shown in table~\ref{tab:5:vals}.
Using the initial values as default, sections \ref{sec:5:max_ddast_threads} to \ref{sec:5:min_ready_tasks} present the results obtained when one of the parameters is modified.
Finally, section~\ref{sec:5:verification} briefly explains how we verified the correctness of \textit{Tuned Value}s.

\subsection{Maximum number of DDAST threads} \label{sec:5:max_ddast_threads}

\begin{figure}[htb!]
   \centering
   \begin{subfigure}[ht]{0.45\linewidth}
      \centering
      \includegraphics[height=0.95\textwidth]{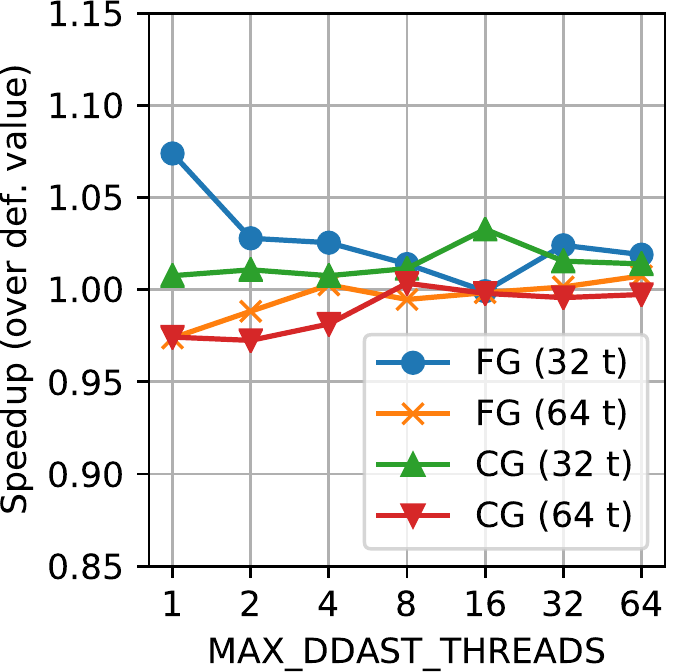}
      \caption{Matmul - KNL} \label{fig:5:max_ddast_threads:dmm:knl}
   \end{subfigure}
   \begin{subfigure}[ht]{0.45\linewidth}
      \centering
      \includegraphics[height=0.95\textwidth]{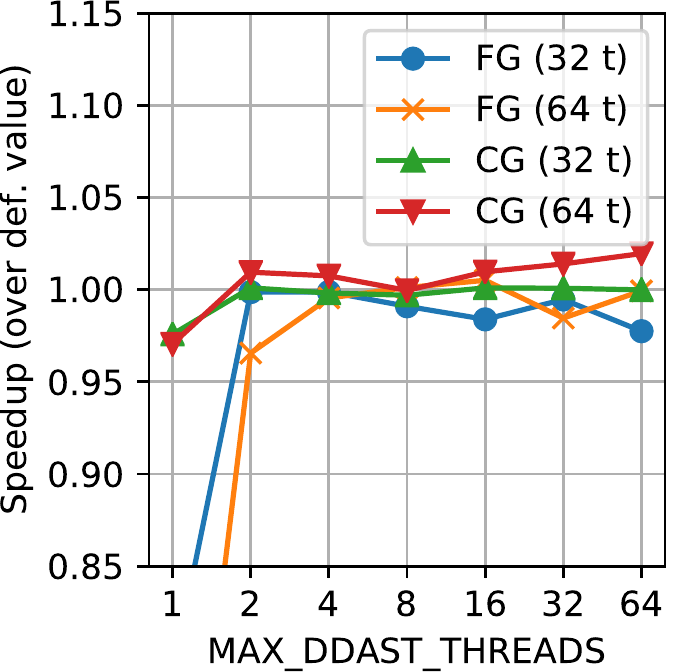}
      \caption{Sparse LU - KNL} \label{fig:5:max_ddast_threads:sparselu:knl}
   \end{subfigure}
   \begin{subfigure}[ht]{0.45\linewidth}
      \centering
      \includegraphics[height=0.95\textwidth]{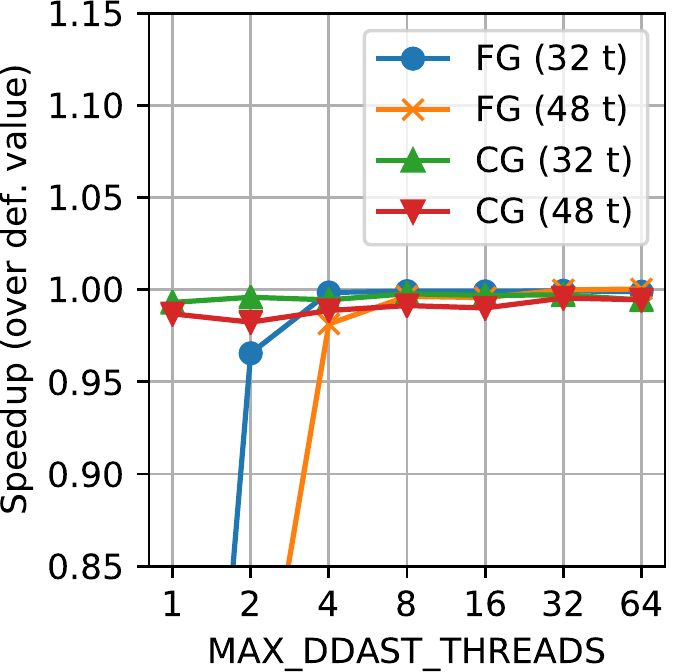}
      \caption{Matmul - ThX} \label{fig:5:max_ddast_threads:dmm:thunderx}
   \end{subfigure}
   \begin{subfigure}[ht]{0.45\linewidth}
      \centering
      \includegraphics[height=0.95\textwidth]{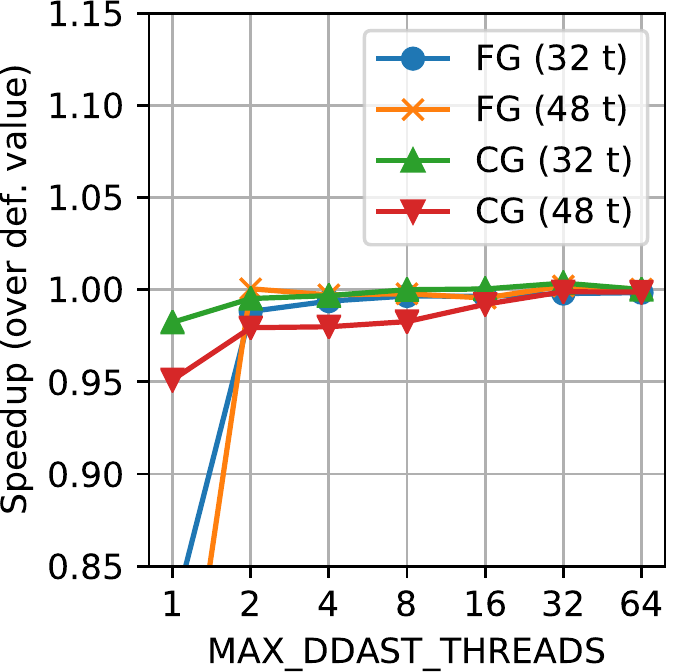}
      \caption{Sparse LU - ThX} \label{fig:5:max_ddast_threads:sparselu:thunderx}
   \end{subfigure}
   \begin{subfigure}[ht]{0.45\linewidth}
      \centering
      \includegraphics[height=0.95\textwidth]{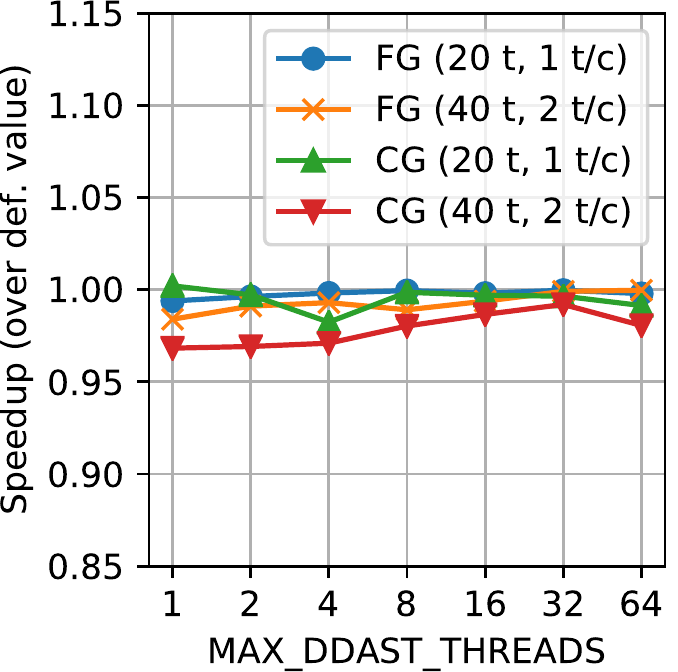}
      \caption{Matmul - Power8+} \label{fig:5:max_ddast_threads:dmm:power}
   \end{subfigure}
   \begin{subfigure}[ht]{0.45\linewidth}
      \centering
      \includegraphics[height=0.95\textwidth]{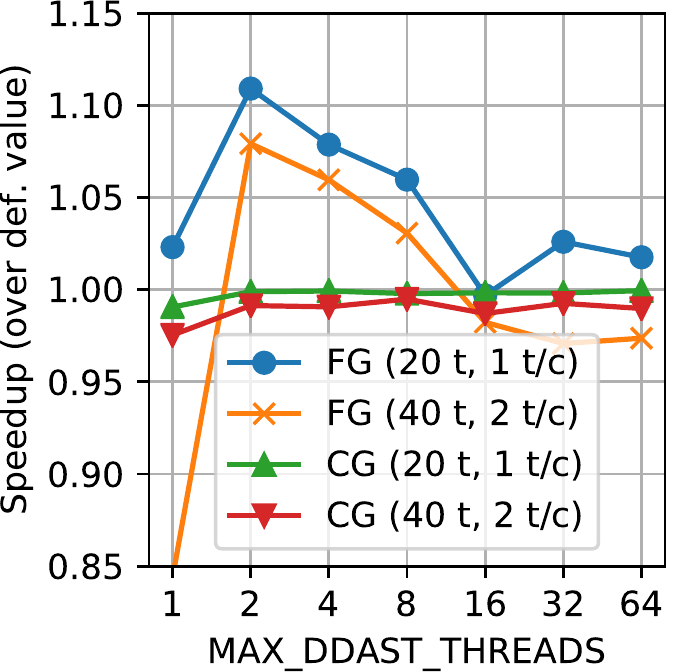}
      \caption{Sparse LU - Power8+} \label{fig:5:max_ddast_threads:sparselu:power}
   \end{subfigure}
   \caption{Speedup changing the \texttt{MAX\_DDAST\_THREADS}} \label{fig:5:max_ddast_threads:dmm}
   \label{fig:5:max_ddast_threads}
\end{figure}

The plots of figure~\ref{fig:5:max_ddast_threads}, one for each benchmark and architecture, show the speedup over the default parameter value (y-axes) when changing the \texttt{MAX\_DDAST\_THREADS} parameter value (x-axes).
The number of threads used in each configuration appears in the legend of the plots.

The results show different behavior depending on the architecture/application combination.
The executions with fine grain tasks are more influenced by the parameter value than coarse grain.
One one hand, some fine grain results (figure~\ref{fig:5:max_ddast_threads:sparselu:knl}, figure~\ref{fig:5:max_ddast_threads:dmm:thunderx} and figure~\ref{fig:5:max_ddast_threads:sparselu:thunderx}) show that only one manager thread may not be able to handle the incoming messages increasing the execution time and limiting the application performance.
In these cases, the results show that the parameter value does not influence the execution time when it goes above 2-4 depending on the number of worker threads.
On the other hand, the SparseLU results in Power8+ show that more than one thread in the DDAST callback is needed to handle the messages but restricting the number of threads to 2-4 may provide a better performance (figure~\ref{fig:5:max_ddast_threads:sparselu:power}).
The improvement comes from the better exploitation of the runtime structures' locality as they are mainly accessed by only 2-4 threads.

Considering all results in figure~\ref{fig:5:max_ddast_threads}, keeping the number of manager threads in the DDAST callback restricted but avoiding creating a bottleneck in the messages processing is the best option.
The smaller the number of manager threads, the better data locality in the execution of runtime operations.
In addition, the number of manager threads must be large enough to process the peaks of messages without harming the application performance.
These peaks are related to the number of worker threads ($num\_threads$), which is known during the runtime initialization, because it shapes the throughput of messages creation.
Therefore, the new default value for \texttt{MAX\_DDAST\_THREADS} based on the number of available worker threads is defined as: \linebreak[4] $\lceil num\_threads/8 \rceil$.

\subsection{Maximum number of spins} \label{sec:5:max_spins}

\begin{figure}[ht!]
   \centering
   \begin{subfigure}[ht]{0.45\linewidth}
      \centering
      \includegraphics[width=0.95\textwidth]{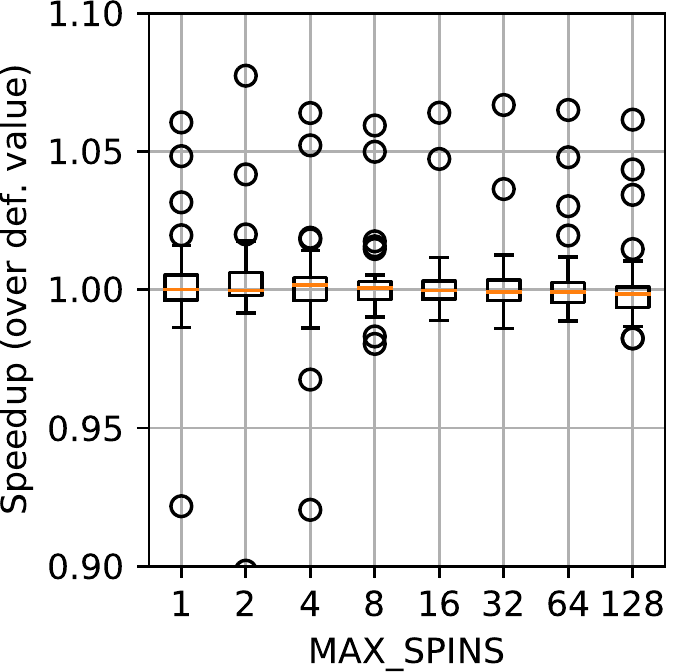}
      \caption{Overall} \label{fig:5:max_spins:global}
   \end{subfigure}
   \begin{subfigure}[ht]{0.45\linewidth}
      \centering
      \includegraphics[width=0.95\textwidth]{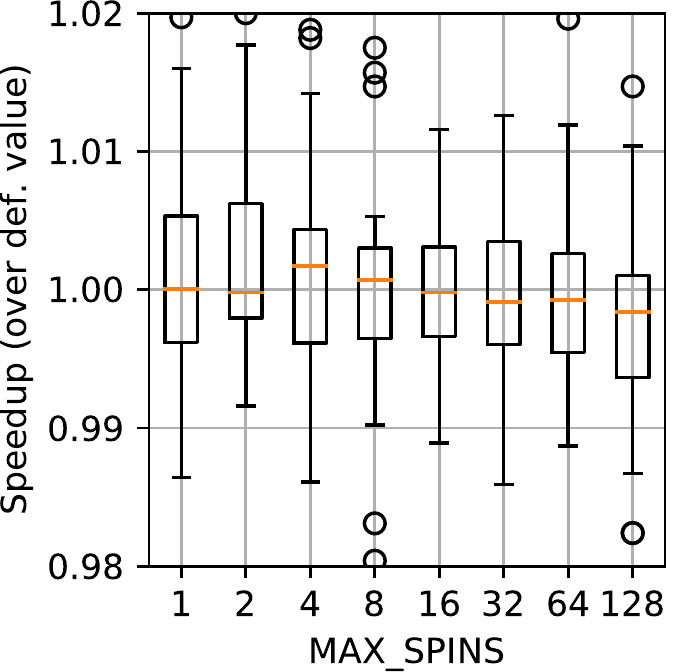}
      \caption{Zoom-in} \label{fig:5:max_spins:fitted}
   \end{subfigure}
   \caption{Speedup changing the \texttt{MAX\_SPINS}} \label{fig:5:max_spins}
\end{figure}

Figure~\ref{fig:5:max_spins} shows the speedup over the default parameter value (y-axes) when changing the \texttt{MAX\_SPINS} parameter value (x-axes).
The figure contains the equivalent results of figure~\ref{fig:5:max_ddast_threads} but summarized in only one plot as all benchmarks and architectures yield very similar results.
Figure~\ref{fig:5:max_spins:global} and figure~\ref{fig:5:max_spins:fitted} contain the same results but using different scales in the y-axes to better observe the behavior.

The results show that the execution time is almost not affected by the value of \texttt{MAX\_SPINS}.
Regardless of the values, the speedup is near to 1 in all the cases and the variability of the results is small ($\pm 0.5\%$) and can be associated with the usual variability between executions rather than the performance difference between the different values.
On one hand, if no other callback functions are registered, the best approach may be use a large value and retain the threads in the DDAST callback until there are some ready tasks.
On the other hand, if there are other registered callback functions, a small parameter value avoids retaining the threads in the DDAST callback until any other break condition in satisfied.
Considering the gathered results and a future scenario where the Functionality Dispatcher is used to manage more runtime functionalities, the new default value for the \texttt{MAX\_SPINS} parameter is set to \textit{1}.

\subsection{Maximum operations per thread} \label{sec:5:max_ops_thread}

\begin{figure}[ht!]
   \centering
   \begin{subfigure}[ht]{0.45\linewidth}
      \centering
      \includegraphics[width=0.95\textwidth]{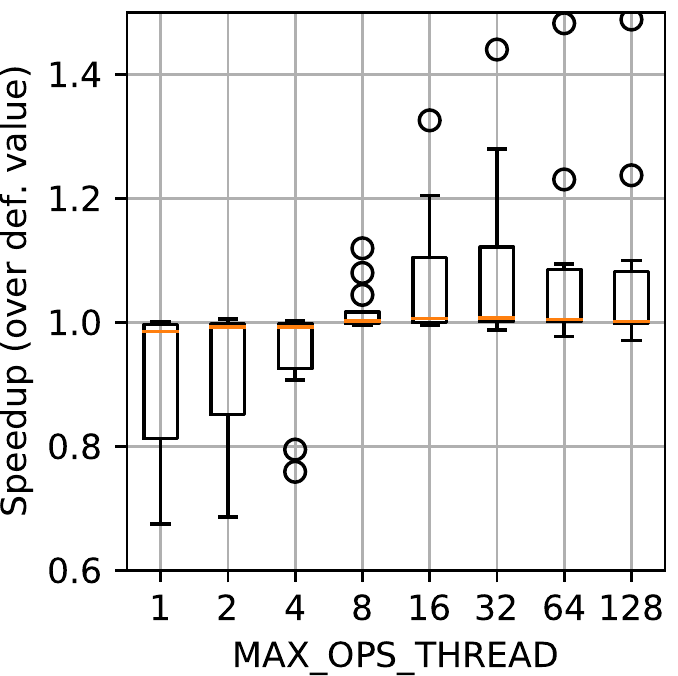}
      \caption{Matmul} \label{fig:5:max_ops_thread:dmm}
   \end{subfigure}
   \begin{subfigure}[ht]{0.45\linewidth}
      \centering
      \includegraphics[width=0.95\textwidth]{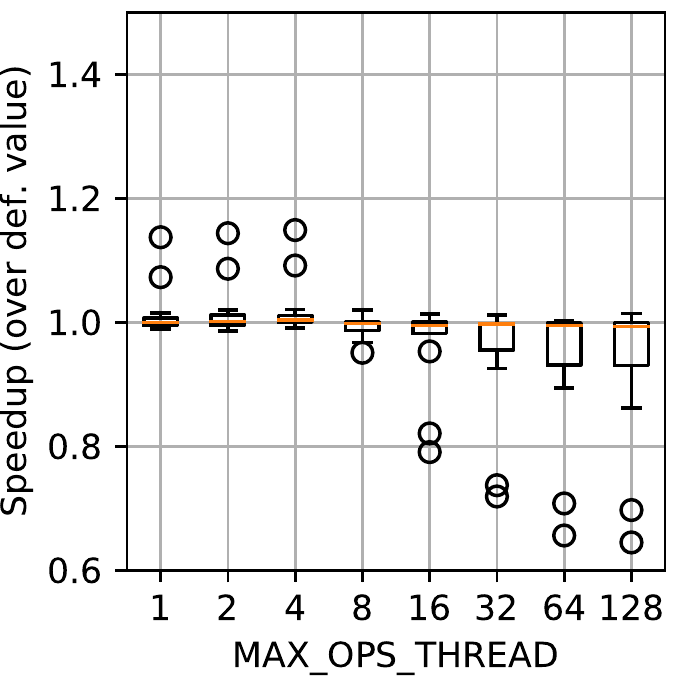}
      \caption{Sparse LU} \label{fig:5:max_ops_thread:sparselu}
   \end{subfigure}
   \caption{Speedup changing the \texttt{MAX\_OPS\_THREAD}} \label{fig:5:max_ops_thread}
\end{figure}

Figure~\ref{fig:5:max_ops_thread} shows the speedup over the default parameter value (y-axes) when changing the \texttt{MAX\_\allowbreak OPS\_\allowbreak THREAD} parameter value (x-axes).
The results are joined by benchmark as each one yields different results, but they behave similarly regardless of the architecture.

The results show that the average speedup for all values in all benchmarks is near to 1 but the variability changes with the parameter value.
On one hand, the Matmul (figure~\ref{fig:5:max_ops_thread:dmm}) have slowdowns when the \texttt{MAX\_OPS\_THREAD} is below 8 and improvements when the value is larger than so.
On the other hand, the Sparse LU benchmark (figure~\ref{fig:5:max_ops_thread:sparselu}) shows opposite behavior with slowdowns when the  \texttt{MAX\_OPS\_THREAD} is above 8.

The opposed behaviors shown are due to the different dependence patterns that each application has.
The execution time may increase with a large value of \texttt{MAX\_OPS\_THREAD} if the ready tasks depend on a \textit{Done Task Message}.
In this case, a critical runtime operation request may be delayed because the DDAST manager will before process a large number of messages from different worker threads .
However, the execution time may decrease with large values of \texttt{MAX\_OPS\_THREAD} if most of the messages trigger the scheduling of a new ready tasks.
In this case, the manager threads may benefit from the data locality when executing the runtime operation request from the same queue.

Considering the results, a reasonable default value for the \texttt{MAX\_OPS\_THREAD} parameter is 8.
This value is the one with less observations below 1 and without any observation which significantly decrease the previous value performance.
In addition, it still has some positive speedup observations.

\subsection{Minimum number of ready tasks} \label{sec:5:min_ready_tasks}

\begin{figure}[ht!]
   \centering
   \begin{subfigure}[ht]{0.45\linewidth}
      \centering
      \includegraphics[width=0.95\textwidth]{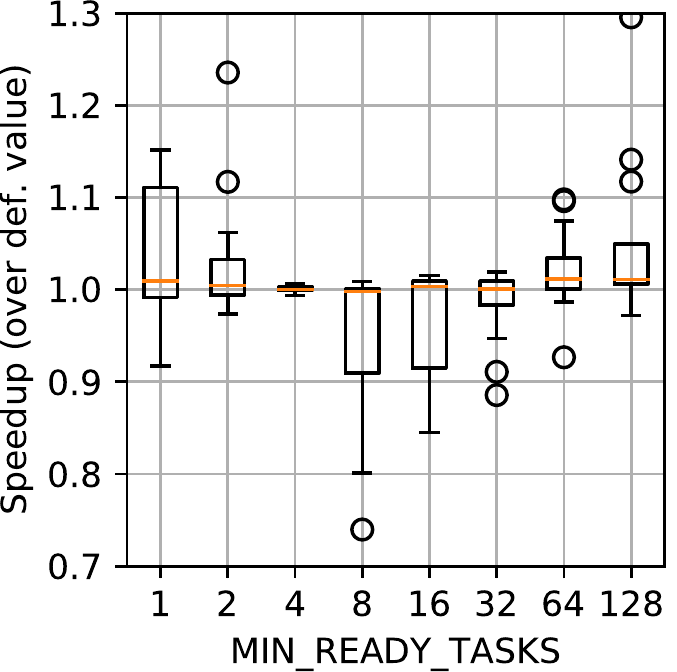}
      \caption{Matmul} \label{fig:5:min_ready_tasks:dmm}
   \end{subfigure}
   \begin{subfigure}[ht]{0.45\linewidth}
      \centering
      \includegraphics[width=0.95\textwidth]{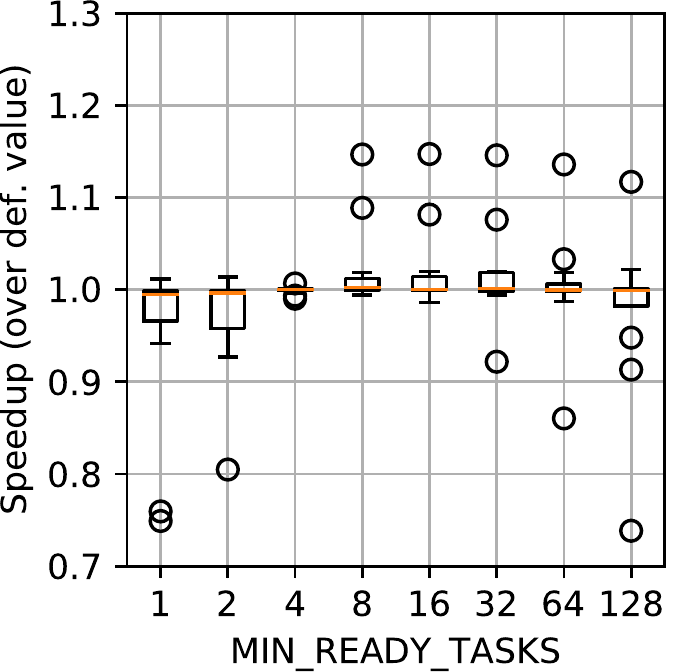}
      \caption{Sparse LU} \label{fig:5:min_ready_tasks:sparselu}
   \end{subfigure}
   \caption{Speedup changing the \texttt{MIN\_READY\_TASKS}} \label{fig:5:min_ready_tasks}
\end{figure}

Figure~\ref{fig:5:min_ready_tasks} shows the speedup over the default parameter value (y-axes) when changing the \texttt{MIN\_\allowbreak READY\_\allowbreak TASKS} parameter value (x-axes).
The results are joined in different plots by benchmark.

The different task dependence patterns result in a different behavior depending on the benchmark, as figure~\ref{fig:5:min_ready_tasks} shows, like for the \texttt{MAX\_OPS\_THREAD} parameter.
However, the average speedup in all applications and values is near to 1 but with different variabilities depending on the parameter value.
In the Matmul benchmark, the values with positive variation and without slowdowns are the more extreme.
The values in the middle pollute the cache to discover some ready tasks but don't benefit from the data locality of loaded information to discover the largest possible number of ready tasks.
In contrast, the SparseLU benchmark has negative variations in the extreme values.
In this benchmark, discovering one ready task is not as easy as in the Matmul so the cache usually becomes polluted.
Moreover, discovering a large number of ready tasks is also complicated, so the manager threads are spinning a lot of time.

Considering all the results, the predefined default value for the \texttt{MIN\_READY\_TASKS} parameter, which is 4, is the best configuration.
The alternatives that outperform this value in some cases are really bad in other ones.

\subsection{Verification} \label{sec:5:verification}

To confirm the results obtained in the previous evaluation, the analysis has been repeated but using the \textit{Tuned Value}s as defaults and also evaluating the N-Body benchmark.
These new results are not shown in the paper as they behave similar to presented ones, that confirms the correctness of \textit{Tuned Value}s defined after the initial analysis.
In addition, the results obtained in this second exploration are used to measure the \textit{DDAST tuned} performance in the next section.

\section{Performance Comparison} \label{sec:6}
The following sections present the performance comparison of the new runtime model against other task-based parallel programming models.
Section~\ref{sec:6:scal} shows the scalability results for the different runtimes and section~\ref{sec:6:anal} presents the most relevant executions where the main differences between both runtime approaches can be seen.

\subsection{Scalability results} \label{sec:6:scal}

\begin{figure}[htb!]
   \centering
   \begin{subfigure}[ht]{0.45\linewidth}
      \centering
      \includegraphics[width=0.95\linewidth]{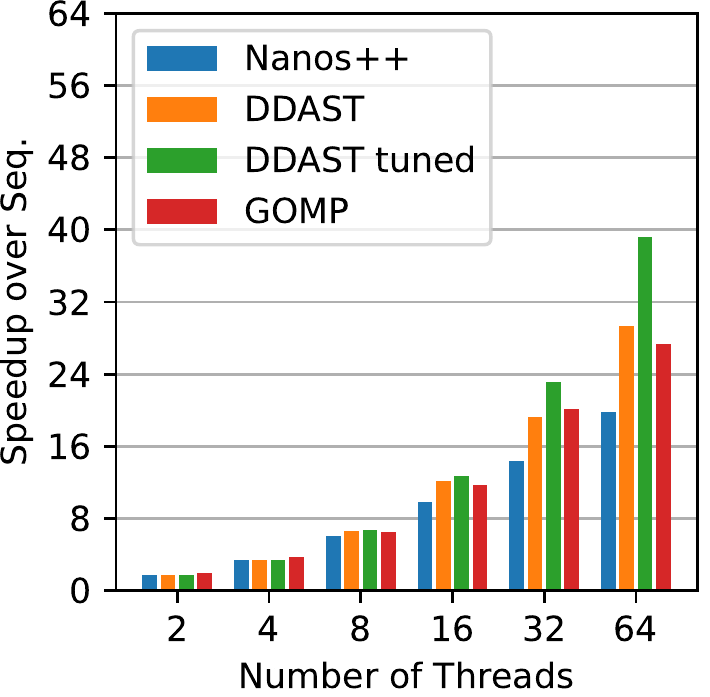}
      \caption{KNL - FG} \label{fig:6:dmm:knl:fine}
   \end{subfigure}
   \begin{subfigure}[ht]{0.45\linewidth}
      \centering
      \includegraphics[width=0.95\linewidth]{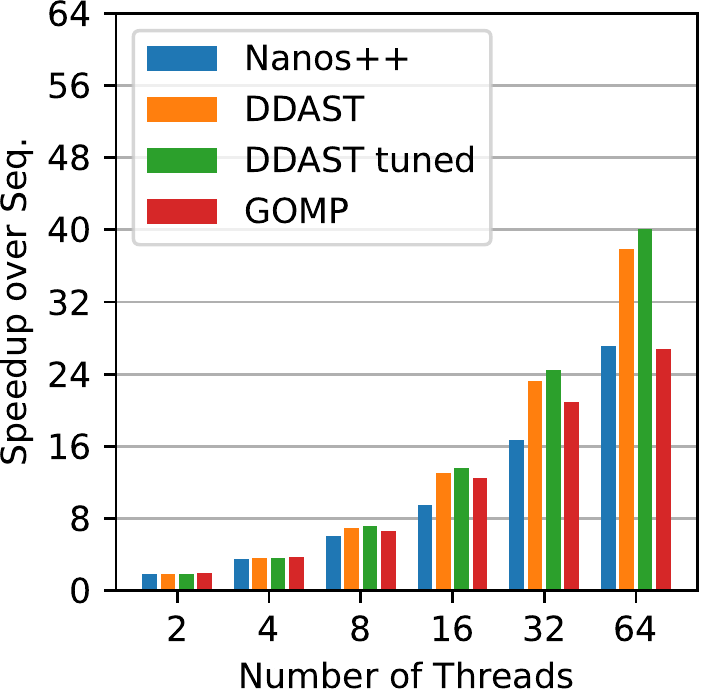}
      \caption{KNL - CG} \label{fig:6:dmm:knl:coarse}
   \end{subfigure}
   \begin{subfigure}[ht]{0.45\linewidth}
      \centering
      \includegraphics[width=0.95\linewidth]{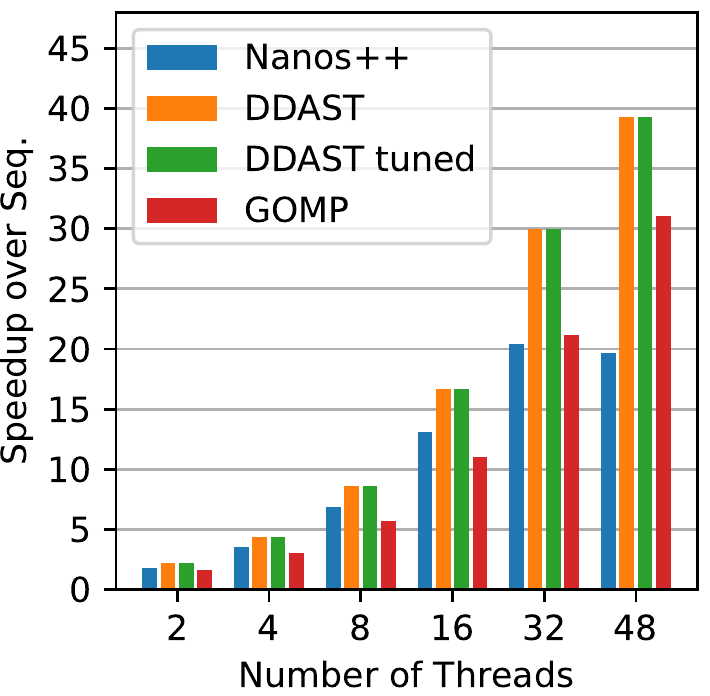}
      \caption{ThunderX - FG} \label{fig:6:dmm:thunderx:fine}
   \end{subfigure}
   \begin{subfigure}[ht]{0.45\linewidth}
      \centering
      \includegraphics[width=0.95\linewidth]{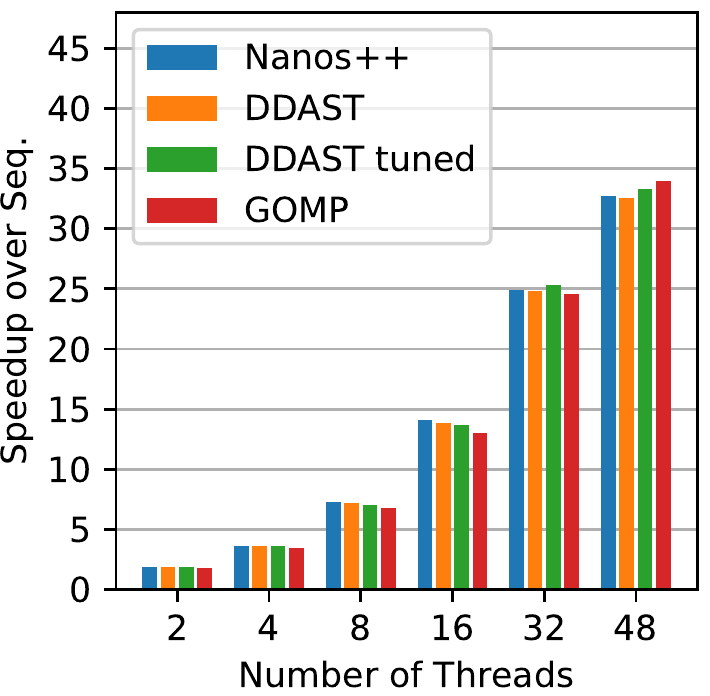}
      \caption{ThunderX - CG} \label{fig:6:dmm:thunderx:coarse}
   \end{subfigure}
   \begin{subfigure}[ht]{0.45\linewidth}
      \centering
      \includegraphics[width=0.95\linewidth]{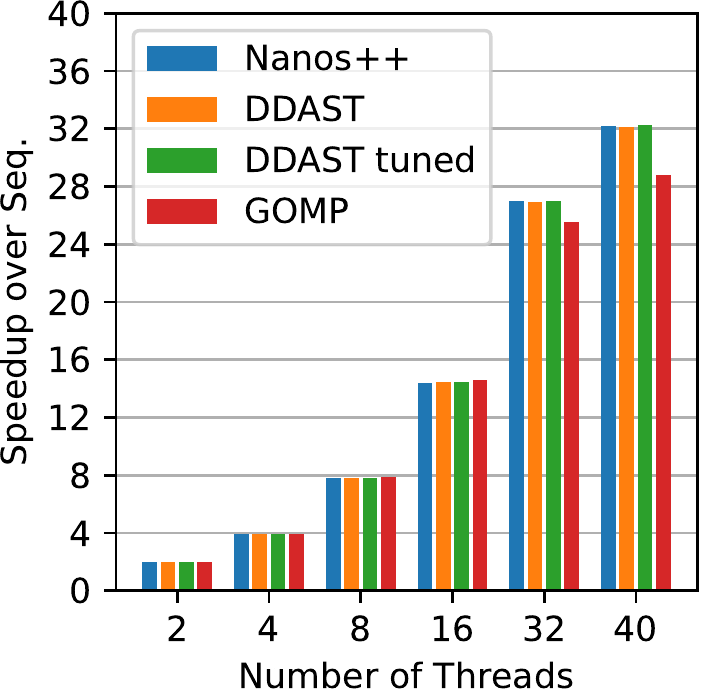}
      \caption{Power9 - FG} \label{fig:6:dmm:pwr:fine}
   \end{subfigure}
   \begin{subfigure}[ht]{0.45\linewidth}
      \centering
      \includegraphics[width=0.95\linewidth]{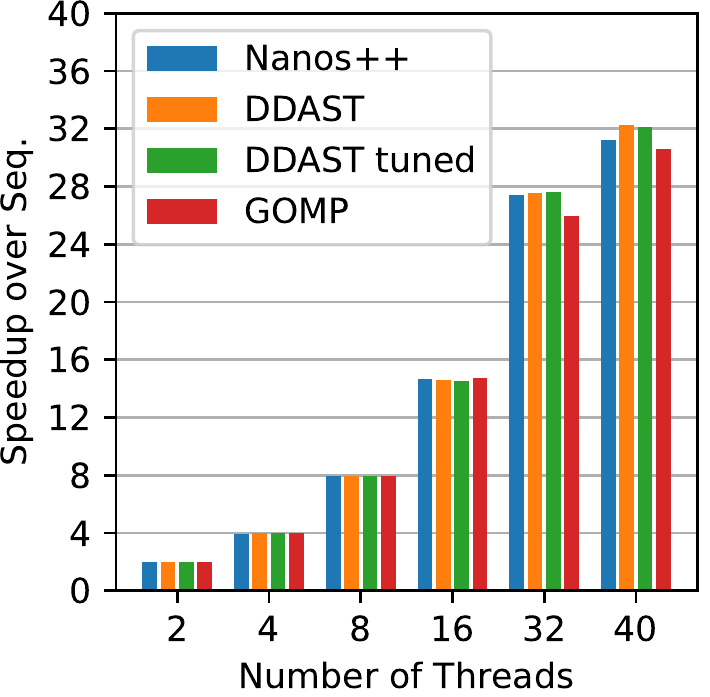}
      \caption{Power9 - CG} \label{fig:6:dmm:pwr:coarse}
   \end{subfigure}
   \caption{Matmul scalability} \label{fig:6:dmm}
\end{figure}

\begin{figure}[htb!]
   \centering
   \begin{subfigure}[ht]{0.45\linewidth}
      \centering
      \includegraphics[width=0.95\linewidth]{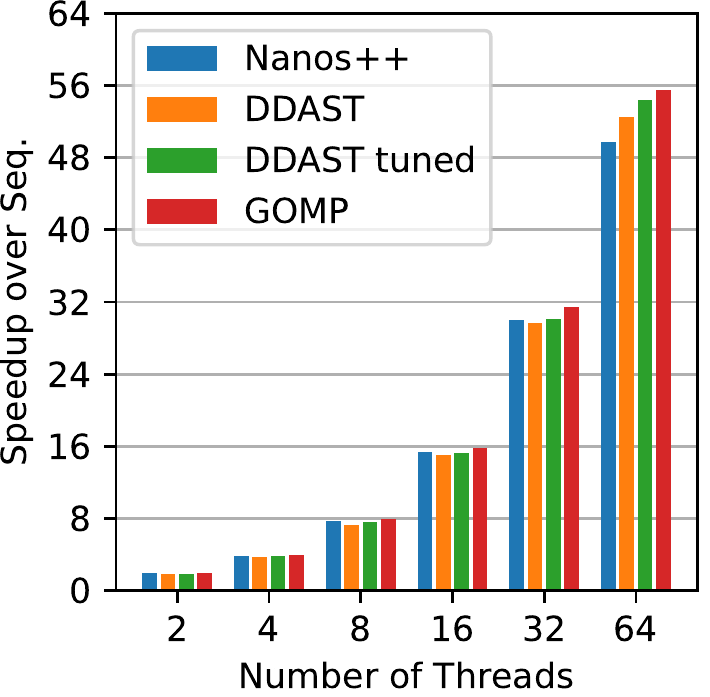}
      \caption{KNL - FG} \label{fig:6:sparselu:knl:fine}
   \end{subfigure}
   \begin{subfigure}[ht]{0.45\linewidth}
      \centering
      \includegraphics[width=0.95\linewidth]{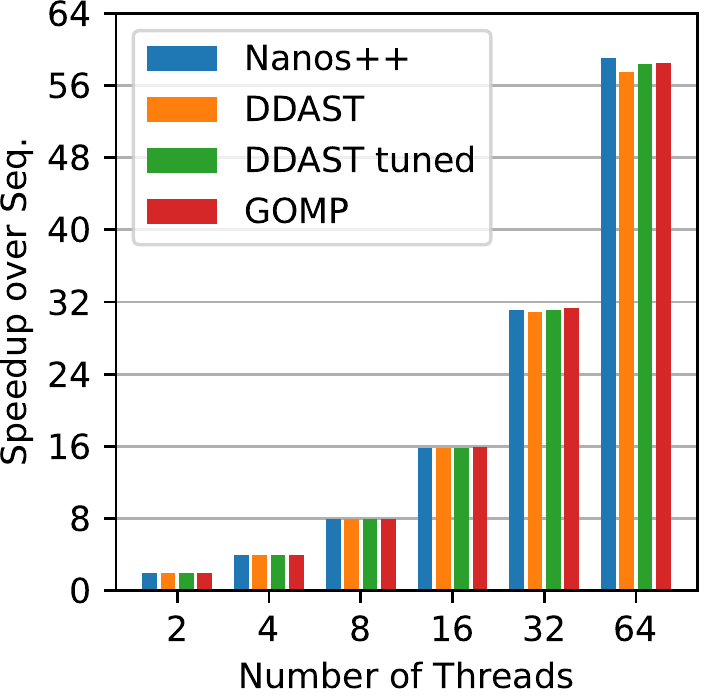}
      \caption{KNL - CG} \label{fig:6:sparselu:knl:coarse}
   \end{subfigure}
   \begin{subfigure}[ht]{0.45\linewidth}
      \centering
      \includegraphics[width=0.95\linewidth]{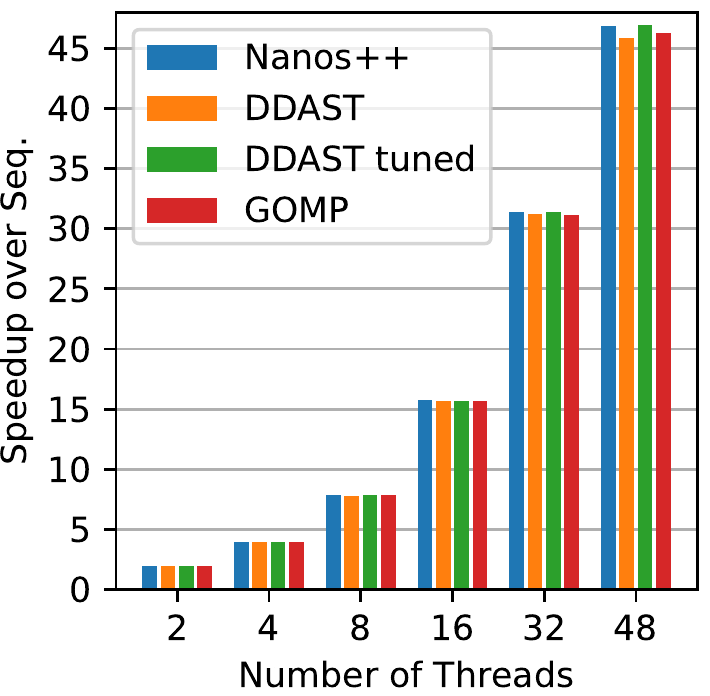}
      \caption{ThunderX - FG} \label{fig:6:sparselu:thunderx:fine}
   \end{subfigure}
   \begin{subfigure}[ht]{0.45\linewidth}
      \centering
      \includegraphics[width=0.95\linewidth]{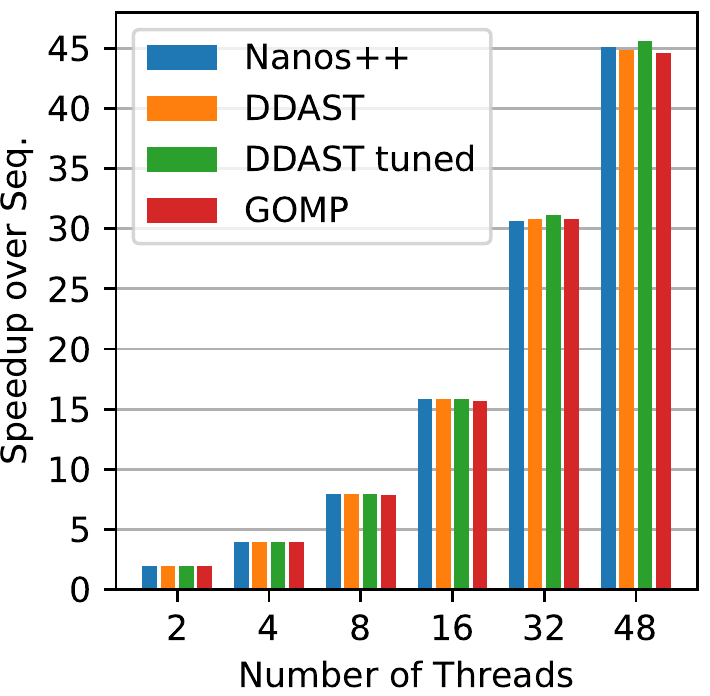}
      \caption{ThunderX - CG} \label{fig:6:sparselu:thunderx:coarse}
   \end{subfigure}
   \begin{subfigure}[ht]{0.45\linewidth}
      \centering
      \includegraphics[width=0.95\linewidth]{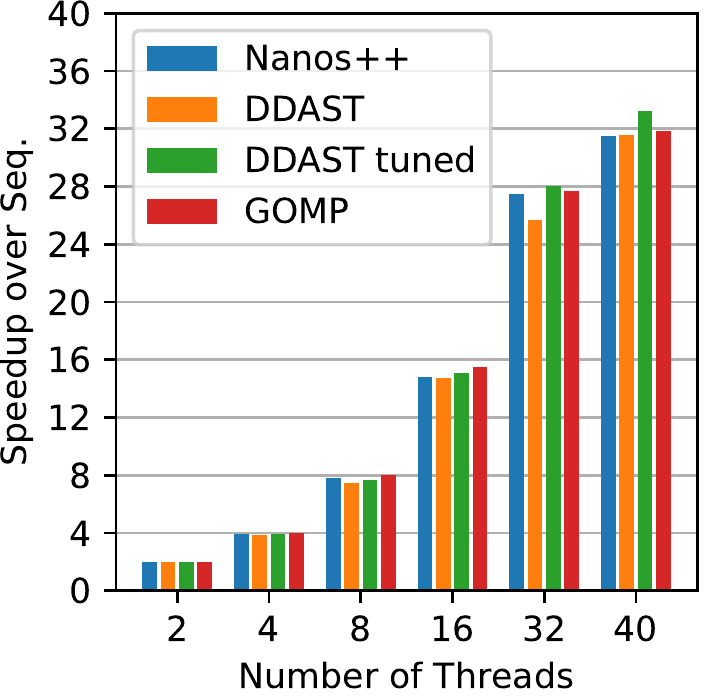}
      \caption{Power9 - FG} \label{fig:6:sparselu:pwr:fine}
   \end{subfigure}
   \begin{subfigure}[ht]{0.45\linewidth}
      \centering
      \includegraphics[width=0.95\linewidth]{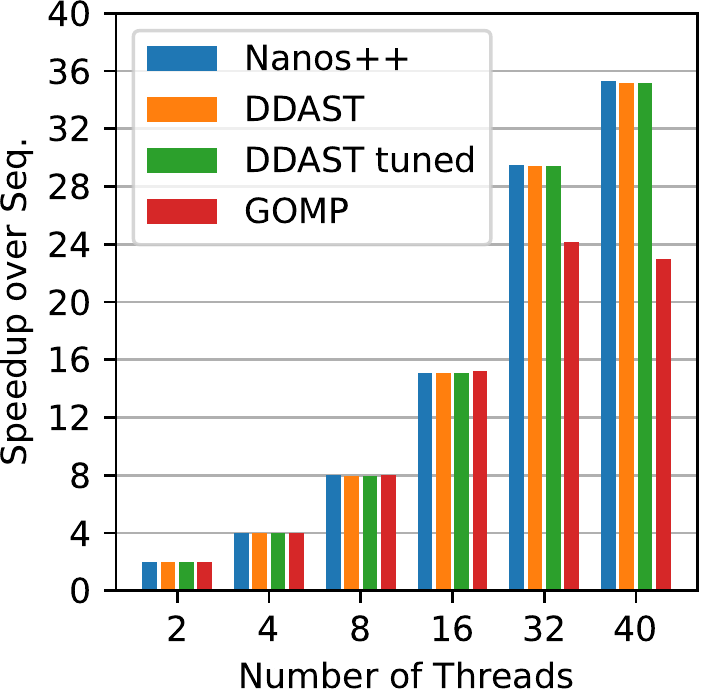}
      \caption{Power9 - CG} \label{fig:6:sparselu:pwr:coarse}
   \end{subfigure}
   \caption{Sparse LU scalability} \label{fig:6:sparselu}
\end{figure}

\begin{figure}[htb!]
   \centering
   \begin{subfigure}[ht]{0.45\linewidth}
      \centering
      \includegraphics[width=0.95\linewidth]{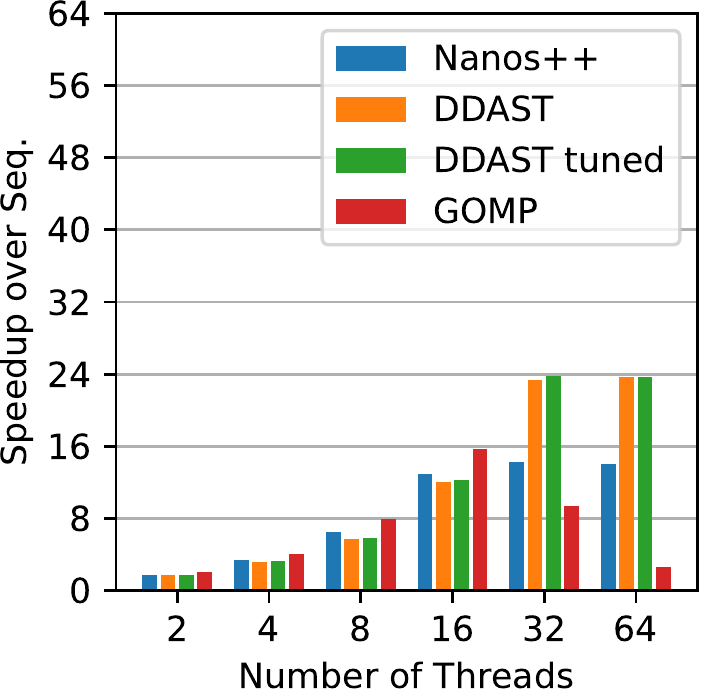}
      \caption{KNL - FG} \label{fig:6:nbody:knl:fine}
   \end{subfigure}
   \begin{subfigure}[ht]{0.45\linewidth}
      \centering
      \includegraphics[width=0.95\linewidth]{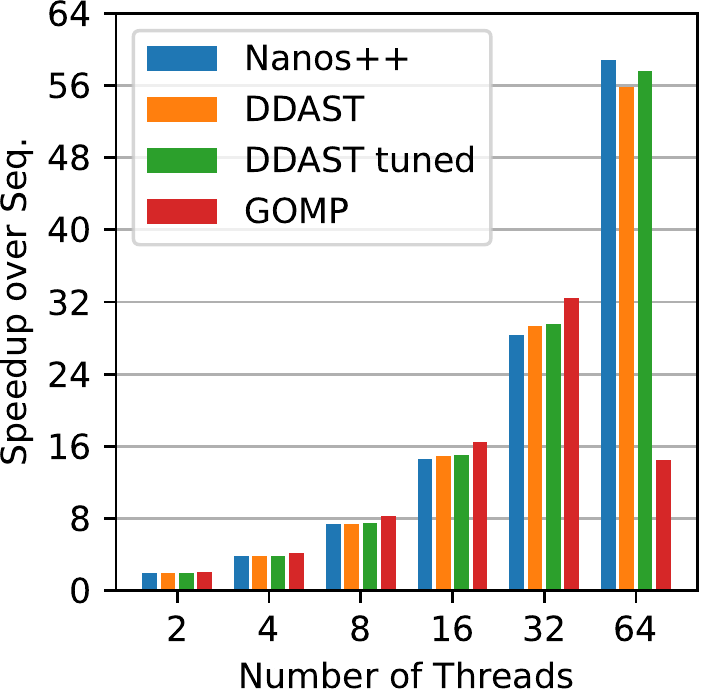}
      \caption{KNL - CG} \label{fig:6:nbody:knl:coarse}
   \end{subfigure}
   \begin{subfigure}[ht]{0.45\linewidth}
      \centering
      \includegraphics[width=0.95\linewidth]{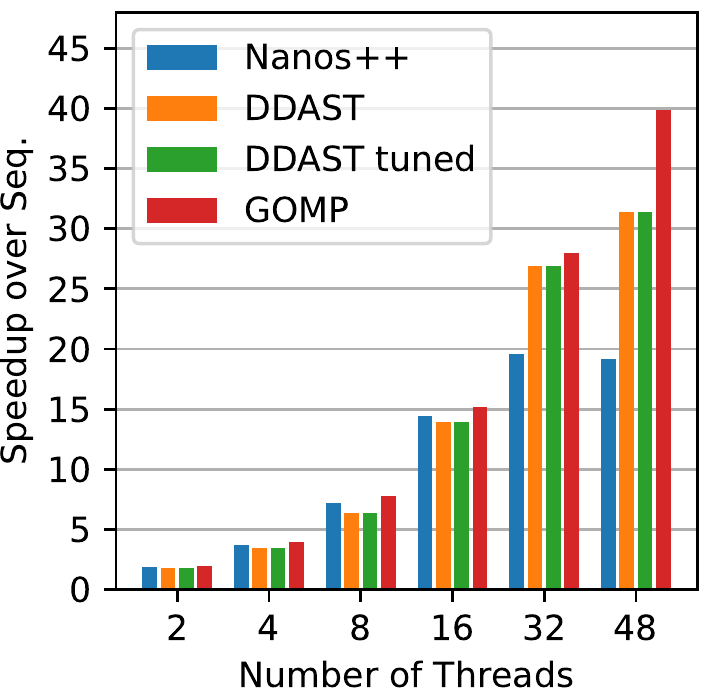}
      \caption{ThunderX - FG} \label{fig:6:nbody:thunderx:fine}
   \end{subfigure}
   \begin{subfigure}[ht]{0.45\linewidth}
      \centering
      \includegraphics[width=0.95\linewidth]{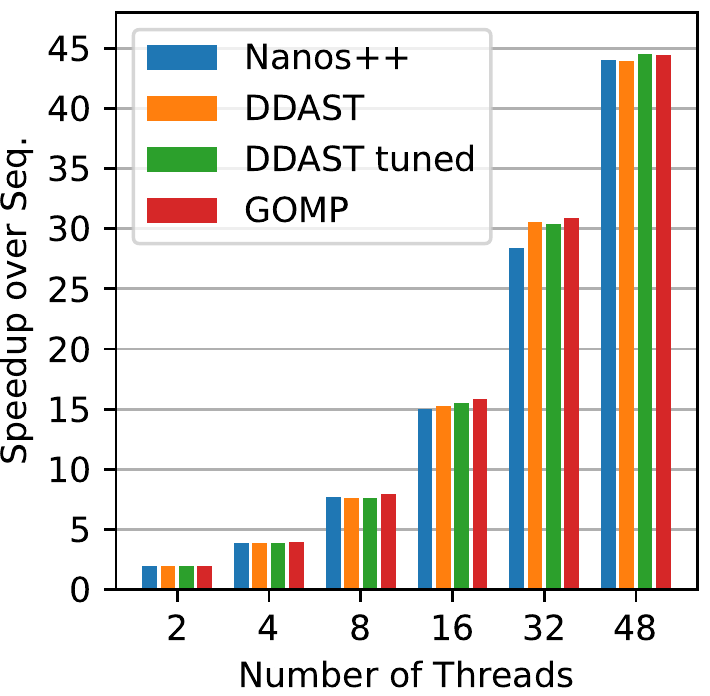}
      \caption{ThunderX - CG} \label{fig:6:nbody:thunderx:coarse}
   \end{subfigure}
   \begin{subfigure}[ht]{0.45\linewidth}
      \centering
      \includegraphics[width=0.95\linewidth]{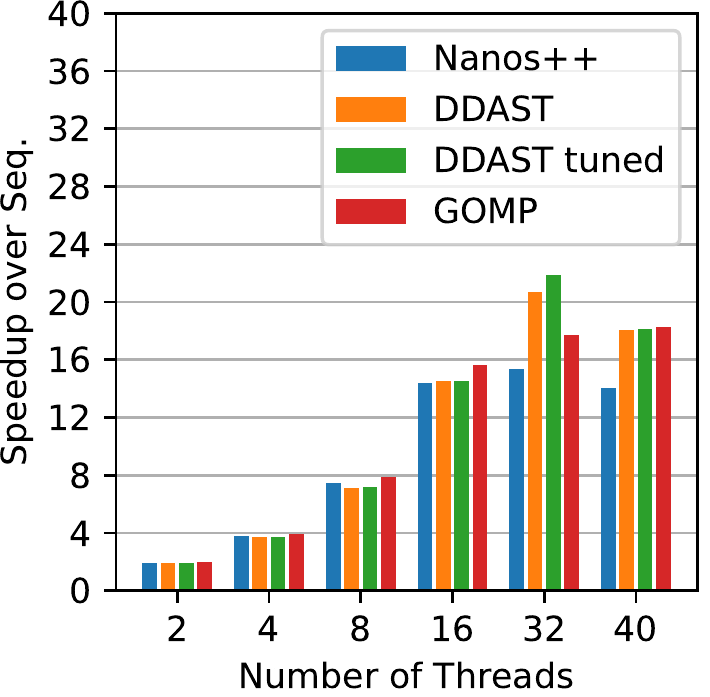}
      \caption{Power9 - FG} \label{fig:6:nbody:pwr:fine}
   \end{subfigure}
   \begin{subfigure}[ht]{0.45\linewidth}
      \centering
      \includegraphics[width=0.95\linewidth]{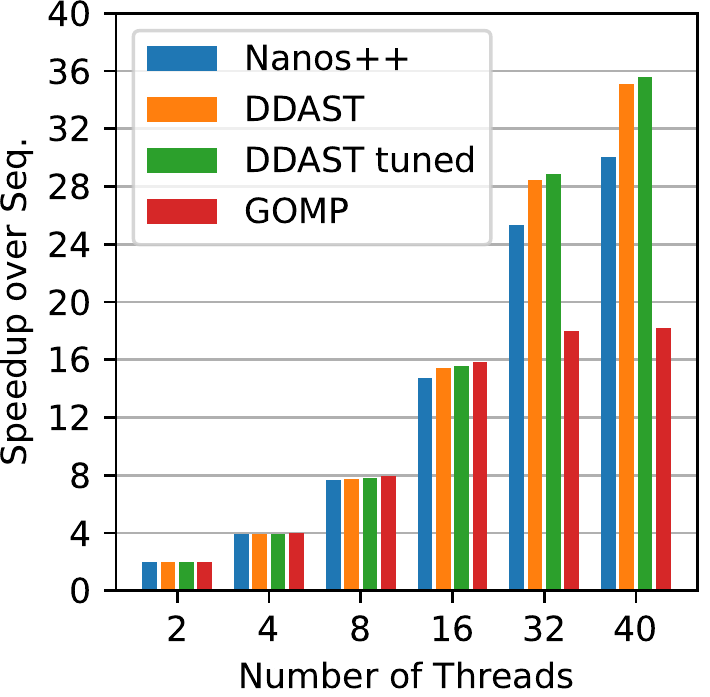}
      \caption{Power9 - CG} \label{fig:6:nbody:pwr:coarse}
   \end{subfigure}
   \caption{N-Body scalability} \label{fig:6:nbody}
\end{figure}

The obtained scalability results are shown in figure~\ref{fig:6:dmm}, figure~\ref{fig:6:sparselu} and figure~\ref{fig:6:nbody} for Matmul, Sparse LU and N-Body respectively.
As we have used Matmul and Sparse LU benchmarks for the tuning, the N-Body results can be used as a control of the performance obtained with a new application.
The results are shown for the KNL, ThunderX and Power9 architectures and for different runtime versions/configurations:

\begin{itemize}
   \item \textit{Nanos++}.
   Baseline OmpSs runtime (version 0.11a).

   \item \textit{DDAST}.
	   Runtime with the distributed runtime manager implementation (DDAST manager) and using the tuned values for the DDAST parameters, which are summarized in table~\ref{tab:5:vals} and are the same for all the runs.
   This version is implemented on top of Nanos++ runtime (version 0.11a).

   \item \textit{DDAST tuned}.
   Same runtime as \textit{DDAST} but with the best values of the DDAST parameters found during the tuning verification for each combination: benchmark, task granularity and architecture.

   \item \textit{GOMP}.
   OpenMP implementation for the GNU Compiler Collection.
   This is a production runtime which performance can be used as a reference of the potential of our approach.
\end{itemize}

The speedup over the sequential version of each benchmark is shown in all plots of figure~\ref{fig:6:dmm}, figure~\ref{fig:6:sparselu} and figure~\ref{fig:6:nbody} (y-axis).
All of them show strong scalability of Nanos++, DDAST and GOMP runtimes for Matmul, Sparse LU and N-Body benchmarks.
Therefore, we can observe the performance evolution when the runtimes must manage more computational resources.
The label of each plot describes the architecture and the task granularity (fine grain, FG, or coarse grain, CG) of those results.
\textit{DDAST tuned} results are included because they show the potential of our proposal.
Also, although it is out of the scope of this work, DDAST manager parameters may be dynamically tuned at runtime to fit each application as shown in \cite{autotune}.

Figure~\ref{fig:6:dmm:knl:fine} and figure~\ref{fig:6:dmm:knl:coarse} show a significant performance improvement ($\sim$40\% for fine grain and $\sim$30\% for coarse grain) when using \textit{DDAST} in comparison to \textit{Nanos++} in KNL architecture for the Matmul benchmark.
The same figures show even a lager improvement for \textit{DDAST tuned} that goes up to $\sim90\%$.

The difference is due to better data locality which reduces the cache misses and consequently the execution time of each task, which is $\sim$33\% shorter in \textit{DDAST} than in \textit{Nanos++}.
The improvement can be explained by the way each approximation accesses the data:
\textit{Nanos++} is accessing the runtime data structures between two task executions, and therefore the runtime is "polluting" the thread caches with its data.
In contrast, \textit{DDAST} avoids those accesses by its asynchronous approach.
The \textit{GOMP} results behave similar to \textit{Nanos++} with better performance in the fine grain results as the GNU runtime has a smaller footprint than \textit{Nanos++}.
The asynchronous approach that drives \textit{DDAST} would also benefit \textit{GOMP} in these situations because the accesses to the shared runtime structures between the execution of tasks will be avoided.

Figure~\ref{fig:6:dmm:thunderx:fine} (Matmul, ThunderX, FG) shows improvement when using \textit{DDAST}, similar to the KNL results.
However, \linebreak[4] \textit{DDAST} gets the same performance as \textit{DDAST tuned} in this architecture.
This means that default values of DDAST manager parameters are the best configuration for such combinations.
Figure~\ref{fig:6:dmm:thunderx:coarse}, figure~\ref{fig:6:dmm:pwr:fine} and figure~\ref{fig:6:dmm:pwr:coarse} show similar results between all runtimes because those configurations do not suffer the contention problem that our proposal tackles.

Regardless of the task granularity, all runtimes provide very good scalability in the Sparse LU benchmark (figure~\ref{fig:6:sparselu}).
The data dependences in this benchmark create an irregular task graph that usually requires processing multiple requests from different worker threads to discover a single ready task.
This creates a challenging situation for the DDAST manager where all possible ready tasks depend on a message which is hidden by several other requests in a queue.
However, the results show that even with this type of applications \textit{DDAST} is able to achieve a performance similar to \textit{Nanos++} and \textit{GOMP}.

Figure~\ref{fig:6:nbody:knl:fine}, figure~\ref{fig:6:nbody:thunderx:fine} and figure~\ref{fig:6:nbody:pwr:fine} (fine grain results of N-Body in the three architectures) show a performance standstill when moving to the larger amount of worker threads for \textit{Nanos++}.
In contrast, \textit{DDAST} maintains or increases the performance compared to the baseline runtime.
The difference between both runtimes is the cost of task submission, which is smaller in \textit{DDAST} due to its asynchronous approach.
This allows the application to create a huge amount of tasks faster in the new runtime model than in the baseline implementation.
In those cases, \textit{GOMP} creates tasks faster than Nanos++ based runtimes for small amounts of worker threads (up to 16 threads in figure~\ref{fig:6:nbody:knl:fine} and up to 32 threads in figure~\ref{fig:6:nbody:knl:coarse}).
However, \textit{GOMP} suffers great contention from the idle worker threads when tasks are executed faster than created, which happens in figure~\ref{fig:6:nbody:knl:fine}  with 32/64 threads and figure~\ref{fig:6:nbody:knl:coarse} with 64 threads.
In ThunderX architecture, \textit{GOMP} does not reach the point where there are several idle worker threads, and consequently, it performs better than both Nanos++ based runtimes.

The coarse grain results in figure~\ref{fig:6:nbody:knl:coarse}, figure~\ref{fig:6:nbody:thunderx:coarse} and figure~\ref{fig:6:nbody:pwr:coarse} do not have such a large performance gap between runtimes as the number of created tasks is a quarter of the amount in fine grain.
Nevertheless, the new runtime organization keeps the performance of the original one.

\subsection{Execution analysis} \label{sec:6:anal}

The Paraver \cite{paraver} execution traces in this section show the behavior of \textit{DDAST} and \textit{Nanos++} among different benchmarks executions to see the main differences between both runtimes.
All execution traces contain the initial and end timestamps in the x-axes (time).
Despite the initial and end timestamps may not match between different runtime executions, any pair of traces that are intended to be compared have the same time duration.
These differences are due to the variable startup overheads that may change between executions, thereby the initial time is adjusted to match the point where the execution of the first task starts.

\begin{figure}[htb!]
   \centering
   \begin{subfigure}[ht]{0.95\linewidth}
      \centering
      \includegraphics[width=\linewidth]{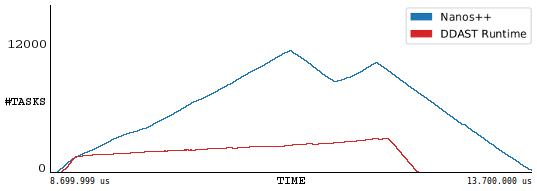}
      \caption{Number of tasks in-graph} \label{fig:6:dmm:knl:fine:traces:in-graph}
   \end{subfigure}
   \begin{subfigure}[ht]{0.95\linewidth}
      \centering
      \includegraphics[width=\linewidth]{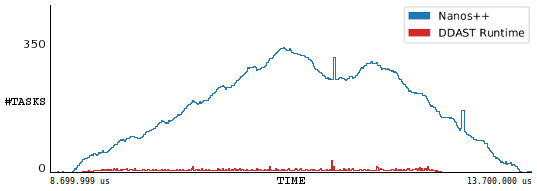}
      \caption{Number of ready tasks} \label{fig:6:dmm:knl:fine:traces:ready}
   \end{subfigure}
   \caption{Execution traces of fine grain Matmul on KNL with 64 threads} \label{fig:6:dmm:knl:fine:traces}
\end{figure}

Figure~\ref{fig:6:dmm:knl:fine:traces} shows the evolution of the number of tasks in the dependence task graph (figure~\ref{fig:6:dmm:knl:fine:traces:in-graph}) and the number of ready tasks (figure~\ref{fig:6:dmm:knl:fine:traces:ready}) along the execution time (x-axis).
Both execution traces are from the same execution of Matmul benchmark in KNL architecture with fine grain tasks.
They have the same duration for the x-axes (5 seconds) but different scales in the y-axes (parameter evolution).
One one hand, \textit{Nanos++} has almost a pyramid shaped evolution where a huge amount of tasks are concurrently managed in the task graph (upper line in figure~\ref{fig:6:dmm:knl:fine:traces:in-graph}) and ready queues (upper line in figure~\ref{fig:6:dmm:knl:fine:traces:ready}).
In fact, the evolution is not a perfect pyramid due to a trace flush to disk at those points, temporally interrupting the main thread and the task creation.
On the other hand, \textit{DDAST} has a roof shaped evolution (bottom lines in figure~\ref{fig:6:dmm:knl:fine:traces:in-graph} and figure~\ref{fig:6:dmm:knl:fine:traces:ready}) where only the minimum amount of tasks needed to discover some parallelism are used, the rest are kept in the manager queues.
This difference greatly influences the runtime overheads which are related to the number of elements that should be managed in the runtime structures.

\begin{figure}[htb!]
   \raggedright
   \begin{subfigure}[ht]{0.95\linewidth}
      \includegraphics[width=\linewidth]{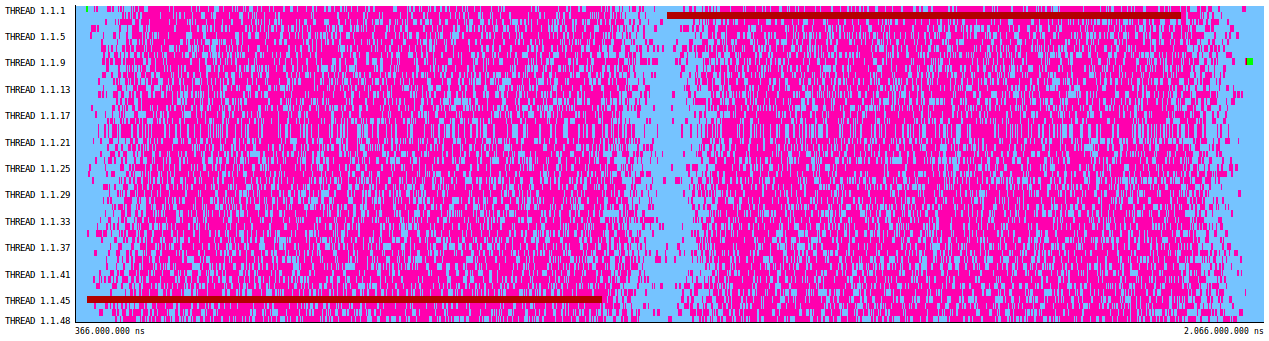}
      \caption{Tasks execution with Nanos++} \label{fig:6:trace:nbody:thunderx:48t:coarse:nanos}
   \end{subfigure}
   \begin{subfigure}[ht]{0.95\linewidth}
      \includegraphics[width=\linewidth]{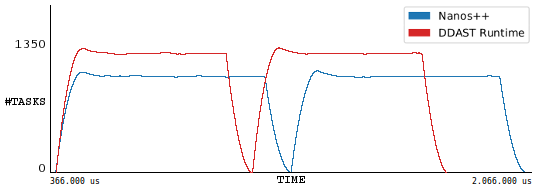}
      \caption{Number of tasks in-graph} \label{fig:6:trace:nbody:thunderx:48t:coarse:in-graph}
   \end{subfigure}
   \begin{subfigure}[ht]{0.95\linewidth}
      \includegraphics[width=\linewidth]{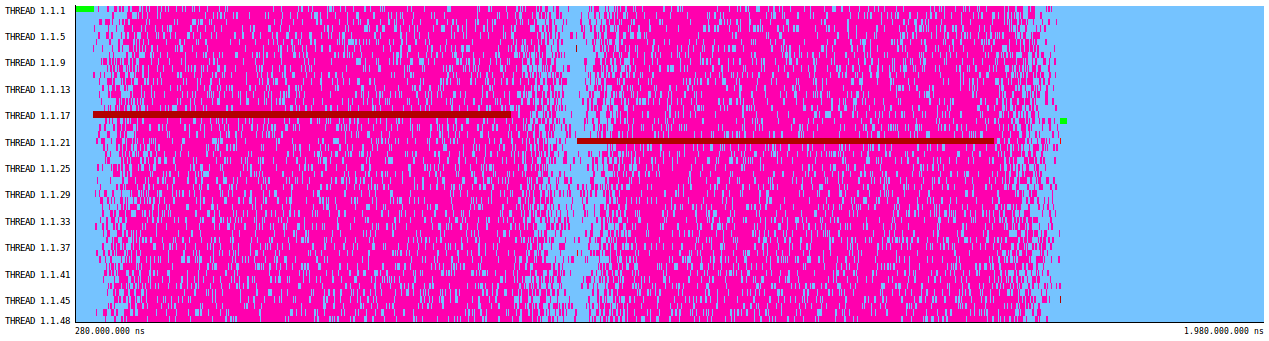}
      \caption{Tasks execution with DDAST} \label{fig:6:trace:nbody:thunderx:48t:coarse:ddast}
   \end{subfigure}
   \caption{Coarse grain N-Body execution traces on ThunderX with 48 threads} \label{fig:6:trace:nbody:thunderx:48t:coarse}
\end{figure}

Figure~\ref{fig:6:trace:nbody:thunderx:48t:coarse} shows different execution traces of N-Body with coarse grain tasks in ThunderX for \textit{Nanos++} and \textit{DDAST}.
However, for clarity purposes, only two timesteps are simulated (instead of the value provided in table~\ref{tab:4:nbody}).
In figures~\ref{fig:6:trace:nbody:thunderx:48t:coarse:nanos} and \ref{fig:6:trace:nbody:thunderx:48t:coarse:ddast}, each line in the y-axes represents one of the 48 worker threads.
In each line, the different colors represent the thread state: sky-blue for IDLE state and other colors for the different task types execution.
Finally, figure~\ref{fig:6:trace:nbody:thunderx:48t:coarse:in-graph} shows the number of tasks in graph (y-axes) evolution for both runtimes.

The tasks execution traces show that the threads execute the tasks at the same throughput that are created.
During each timestep, the execution of the top level task (two long dark-blue regions in figure~\ref{fig:6:trace:nbody:thunderx:48t:coarse:nanos}) ends almost at the same time as children tasks (small pink regions), which are created by the first one.
In this case, the number of ready tasks is near to zero all the time.
Therefore, the faster the task creation, the better performance and resource usage.
So, the reduction in the execution time is from the faster execution of the top level tasks and not from a shorter execution time of leaf tasks like in the Matmul case.

In \textit{DDAST}, the requests to the runtime manager are quickly processed by the manager because the worker threads become idle very frequently.
This increases the throughput of task submission in comparison the the baseline runtime, as the same amount of tasks is created with less time.
Therefore, the number of tasks in the dependence graph increases as can be seen in figure~\ref{fig:6:trace:nbody:thunderx:48t:coarse:in-graph}.
In addition, the same behavior has been observed in the fine grain executions of the same benchmark explaining the better performance results of \textit{DDAST} against \textit{Nanos++} results.

\begin{figure}[htb!]
   \centering
   \begin{subfigure}[ht]{0.95\linewidth}
      \centering
      \includegraphics[width=\linewidth]{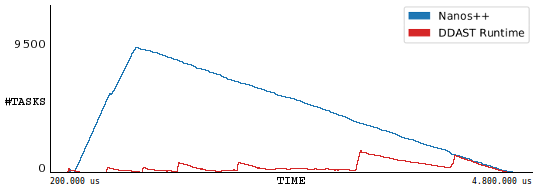}
      \caption{Number of tasks in-graph} \label{fig:6:sparselu:thunderx:coarse:traces:in-graph}
   \end{subfigure}
   \begin{subfigure}[ht]{0.95\linewidth}
      \centering
      \includegraphics[width=\linewidth]{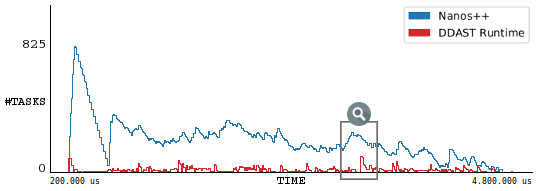}
      \caption{Number of ready tasks} \label{fig:6:sparselu:thunderx:coarse:traces:ready}
   \end{subfigure}
   \caption{Execution traces of coarse grain Sparse LU on ThunderX with 48 threads} \label{fig:6:sparselu:thunderx:coarse:traces}
\end{figure}

The evolution among all the execution time of Sparse LU in ThunderX for the number of tasks in-graph and ready tasks is shown in figure~\ref{fig:6:sparselu:thunderx:coarse:traces}.
Each figure contains a trace for \textit{Nanos++} and \textit{DDAST} runtimes for the entire execution.
The evolution of those parameters is equivalent to the observed in the Matmul benchmark.
The number of in-graph tasks has a pyramid shaped evolution in \textit{Nanos++} and a plain shape with small peaks in \textit{DDAST}.

\begin{figure}[ht]
   \centering
   \begin{subfigure}[ht]{0.95\linewidth}
      \centering
      \includegraphics[width=\linewidth]{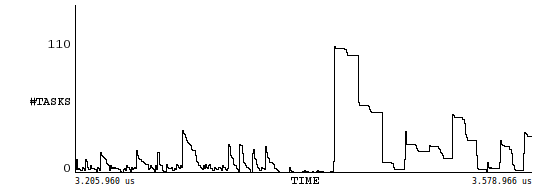}
      \caption{Number of ready tasks} \label{fig:6:zoom:ddast:sparselu:thunderx:48t:coarse:ready}
   \end{subfigure}
   \begin{subfigure}[ht]{0.95\linewidth}
      \centering
      \includegraphics[width=\linewidth]{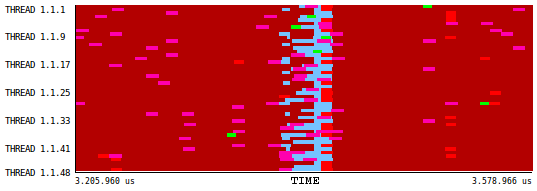}
      \caption{Tasks execution} \label{fig:6:zoom:ddast:sparselu:thunderx:48t:coarse:trace}
   \end{subfigure}
   \caption{Coarse grain Sparse LU partial execution traces on ThunderX with 48 threads and DDAST} \label{fig:6:zoom:ddast:sparselu:thunderx:48t:coarse}
\end{figure}

Figure~\ref{fig:6:zoom:ddast:sparselu:thunderx:48t:coarse} shows two execution traces when almost all threads become idle for the small portion marked in figure~\ref{fig:6:sparselu:thunderx:coarse:traces:ready}.
Figure~\ref{fig:6:zoom:ddast:sparselu:thunderx:48t:coarse:trace} shows the tasks that are executed in each thread and figure~\ref{fig:6:zoom:ddast:sparselu:thunderx:48t:coarse:ready} shows the number of ready tasks (y-axis), both among the execution time (x-axes).
Figure~\ref{fig:6:zoom:ddast:sparselu:thunderx:48t:coarse:ready} shows that the number of ready tasks becomes nearly zero for a relatively long portion of time.
Therefore, almost all \linebreak[4] threads are idle as can be seen in figure~\ref{fig:6:zoom:ddast:sparselu:thunderx:48t:coarse:trace}, so they start to process the pending requests to the runtime manager.
In addition, several tasks are added to the task dependence graph but their data dependences are not satisfied so the tasks do not become ready.
They do so when the \textit{Task Finalization} requests of the critical tasks are processed, at this point the number of ready tasks suddenly increases from zero to more than 100 as can be seen in figure~\ref{fig:6:zoom:ddast:sparselu:thunderx:48t:coarse:ready}.

\section{Related Work} \label{sec:7}
Several works exist about characterization and improvement of parallel programming models.
They are over different models working at different levels and with different approaches.
OmpSs tools (Mercurium and Nanos++), which are open source and are the ones used to test our model, can execute inter-node and intra-node applications \cite{ompssCluster} and are under constant development introducing new features.
Moreover, several people use this programming model as a base to develop different prototypes or extend its functionality.

Previous works discussed the task scheduling and dependences resolution overheads in data-driven task-based models like OpenMP and OmpSs.
TurboBLYSK \cite{Podobas2014} is a framework which implements the OpenMP 4.0 with a custom compiler and a highly efficient runtime schedule of tasks with explicit data-dependence annotations.
Its objective is also to reduce the dependence management overheads of the runtime.
However, TurboBLYSK approach requires extra information in the task dependences definition to allow the runtime to re-use previously resolved dependence patterns and to reduce the overall overhead.
In contrast, our proposal only uses the information provided by default to reduce the overall task management overhead, so our optimization is transparent to the programmers.

Other task-based programming models like Intel Threading Building Blocks \cite{itbb} and Charm++ \cite{charm++} use a execution model that is more pure data-flow than the OmpSs/OpenMP model which has a hybrid (control/data flow) model \cite{yazdanpanah2014hybrid}.
This execution model usually allows to exploit better the parallelism of the applications but requires a specific structure and an application redesign.
In the context of the Intel Threading Building Blocks, there is a previous work discussing the cost of the synchronization inside its runtime, but they focus the problem in the work distribution \cite{improving_itbb} instead of the task graph management that is implicitly done in their execution model.
The Charm++ programming model is intended to provide some valuable features for executions in large computation systems like migratability, checkpoint application restarting, process failure tolerance, malleability, etc.
They have previous work about optimizing the communications inside their runtime \cite{improving_itbb}, but the dataflow model that they have moves the complexity of task graph management into the application development process like in TBB.

HPX (High Performance ParalleX) \cite{hpx} and STAPL (Standard Template Adaptive Parallel Library) \cite{stapl} are general purpose frameworks for parallel and distributed applications of any scale.
Both use the same asynchronous philosophy that we have used in the DDAST design to avoid starvation and reduce the runtime overheads.
However, they require changes in the application implementation to make use of the library.
Also, there are OpenMP implementations over HPX: hpxMP \cite{hpxmp}, which has assembly parts only developed for x86 architectures; and OMPX \cite{ompx}.
The OMPX implementation and our work achieve a similar asynchronous task management.
They use the \texttt{async} and \texttt{future} constructs to achieve the asynchronous task management.
In contrast, we implemented the asynchronous task management in the runtime core.
That allows us to parametrize and control where the runtime management happens.

The DAGuE framework \cite{dague} offers an architecture aware scheduling and management of micro-tasks on distributed many-core heterogeneous architectures.
It uses a mixed control-data flow similar to OmpSs programming model and it also optimizes the tasks management minimizing the amount of tasks created in the system during the execution.
The main difference with our approach is that DAGuE auto-parallelizes the application based on an static analysis of the application at compile time, instead of dynamically building a task dependence graph at runtime.
This difference reduces the amount of applications that can use DAGuE as not all of them have a static dependence pattern that can be formally expressed at compile time.

Other works, which also try to accelerate current runtimes, propose moving part of the runtime into a specific hardware of FPGAs.
Some examples are Nexus\# \cite{nexussharp} and Picos \cite{picos}.
They present different hardware designs that can manage tasks dependences of task-based programming models.
Besides these, there is active research in new computer architectures able to manage efficiently tasks in StarSs family.
For example, some research aims to look for a new Runtime-Aware Architecture to overcome current multi-core restrictions like power, programmability and resilience \cite{romolPaper}.
The main difference between those works and the one proposed in this project is the way to improve the existing system.
They proposed new hardware to work in harmony with the software in order to improve the performance.
In contrast, this project improves the existing parallel programming model runtimes with software ideas that do not require additional hardware.

\section{Conclusions} \label{sec:8}
The multicore processors have become popular and are pre\-sent in almost any electronic device nowadays.
Task-based parallel programming models, like OmpSs, facilitate programmers to use such processor architectures by simply annotating the sequential applications source code.
However, the runtime libraries that support such models present a contention problem when the number of threads grows to some tens.
As current many-core processors, the future processors are expected to have several cores; thereby the runtimes may become a performance bottleneck.

This paper presents a design of an asynchronous runtime structure based on a distributed runtime manager.
It is based on requests from the worker threads to the runtime manager, which modifies the runtime structures handling the requests.
The distributed manager design is based on the idea that any thread should be allowed to become a manager thread if needed.
Despite the fact that we implemented the design over one runtime, it can be applied to any task based runtime system like any of the OpenMP compliant runtimes.
This is because the task life cycle should be really close in all of them and we just change the transition and management of those states.
Even more, the design could be adapted for particular heterogeneous architectures, like big.LITTLE, allowing a subset of the worker threads to become manager threads.

As a proof-of-concept, the paper also presents an implementation of the asynchronous runtime design based on the Nanos++ runtime library.
In such implementation, the runtime core is extended with new generic modules that provide the possibility of asynchronously executing other runtime services by idle worker threads.
These new modules could be used for other runtime actions like sending tasks to accelerators or processing the finished ones.
Finally, the performance results show that the current implementation outperforms the baseline runtime for large amounts of threads and that it has the same performance for small amounts of threads or when the tackled problem does not arise.
As a future work, the runtime manager will dynamically tune its parameters to fit the application requirements and increase even more the performance of our proposal.

\section*{Acknowledgements}
This work is partially supported by the European Union H2020 Research and Innovation Action (projects 801051, 754337 and 780681), by the Spanish Government (projects SEV-2015-0493 and TIN2015-65316-P, grant BES-2016-078046), and by the Generalitat de Catalunya (contracts 2017-SGR-1414 and 2017-SGR-1328).

\bibliographystyle{elsarticle-num}
\bibliography{bibliography.bib}

\end{document}